\documentclass[%
superscriptaddress,
preprint,
nobibnotes,
 amsmath,amssymb,
 aps,
floatfix,
]{revtex4-2}

\usepackage{graphicx}% Include figure files
\usepackage{dcolumn}% Align table columns on decimal point
\usepackage{bm}% bold math
\begin{document}

% \preprint{APS/123-QED}

\title{Structural Signatures for Thermodynamic Stability in Vitreous Silica: Insight from Machine Learning and Molecular Dynamics Simulations}% Force line breaks with \\

\author{Zheng Yu}
\affiliation{Department of Materials Science and Engineering, University of Wisconsin-Madison, Madison 53706, United States}

\author{Qitong Liu}
\affiliation{Department of Civil and Environmental Engineering, University of Wisconsin-Madison, Madison 53706, United States}

\author{Izabela Szlufarska}
\affiliation{Department of Materials Science and Engineering, University of Wisconsin-Madison, Madison 53706, United States}

\author{Bu Wang}%
 \email{bu.wang@wisc.edu}
\affiliation{Department of Materials Science and Engineering, University of Wisconsin-Madison, Madison 53706, United States}
\affiliation{Department of Civil and Environmental Engineering, University of Wisconsin-Madison, Madison 53706, United States}%


\begin{abstract}
The structure-thermodynamic stability relationship in vitreous silica is investigated using machine learning and a library of 24,157 inherent structures generated from melt-quenching and replica exchange molecular dynamics simulations. We find the thermodynamic stability, i.e., enthalpy of the inherent structure ($e_{\mathrm{IS}}$), can be accurately predicted by both linear and nonlinear machine learning models from numeric structural descriptors commonly used to characterize disordered structures. We find short-range features become less indicative of thermodynamic stability below the fragile-to-strong transition. On the other hand, medium-range features, especially those between 2.8--$\sim$6 {\AA}, show consistent correlations with $e_{\mathrm{IS}}$ across the liquid and glass regions, and are found to be the most critical to stability prediction among features from different length scales. Based on the machine learning models, a set of five structural features that are the most predictive of the silica glass stability is identified. 
% \begin{description}
% Secondary publications and information retrieval purposes.
% You may use the \texttt{description} environment to structure your abstract;
% use the optional argument of the \verb+\item+ command to give the category of each item. 
% \end{description}
\end{abstract}

\maketitle

%\tableofcontents

\section{Introduction}

%%\paragraph 
As a non-equilibrium material, glass of a given composition can exhibit different thermodynamic stabilities depending on its thermal history.\cite{debenedettiSupercooledLiquidsGlass2001a,guptaBasisGlassStates2019} The thermodynamic stability, which describes the state of the glass on its potential energy landscape (PEL), affects various static and dynamic properties, e.g., mechanical modulus, internal friction, and kinetic stability.\cite{sundararamanMechanicalPropertiesSilica2016,queenExcessSpecificHeat2013,raegenUltrastableMonodispersePolymer2020} In an effort to modify the glass properties, traditional methods such as annealing or quenching at different rates can be utilized to manipulate the state of the glass. Moreover, it has been demonstrated that the thermodynamic stability of vapor deposited glass can be tailored to a greater extent by tuning the deposition conditions. This has led to ultrastable glasses with unusual behaviors and holds promise for fabricating glasses with exceptional properties.\cite{kearnsInfluenceSubstrateTemperature2007,singhUltrastableGlassesSilico2013a,rafols-ribeHighperformanceOrganicLightemitting2018} 

The ability for glass to exhibit different thermodynamic stabilities originates from its disordered atomic structure. It is generally believed that, as glass becomes more stable, some degree of ordering is enhanced within its structure, e.g., enhanced ordering at short or medium ranges. However, even for some of the most common glass systems like silica, the structure-stability relationship is not well understood. The nature of ordering in glass is usually complicated and could manifest through a range of structural features. Some of these are common features shared with other material families, e.g., density, bond length and angle, and coordination number, while others are more complex. For instance, the first sharp diffraction peak (FSDP) is an universal feature in the structure factor of tetrahedral glass formers and is related to their medium range order. The corresponding real-space features, however, are still debated.\cite{elliottOriginFirstSharp1991,duFirstSharpDiffraction2005,shiDistinctSignatureLocal2019} While it is common to observe multiple structural features change simultaneously with stability,\cite{ritlandDensityPhenomenaTransformation1954,bruningVolumeMathrmBMathrmO1994,foxMolecularDynamicsSimulations1984,vollmayrCoolingrateEffectsAmorphous1996} unknown correlations that might exist between features make identifying the controlling mechanism difficult. In addition, features of a disordered structure are statistical in nature. Whether simple statistical properties of these features, such as the average or the variation, can sufficiently capture the structure-stability relationship remains unclear. 

In this study, we explore machine learning methods to demonstrate the possibility of predicting thermodynamic stability from common structural features characterizing glassy structures, using model silica based on the BKS potential as an example. The BKS silica is a well studied glassy system that bears a close resemblance to the real silica system. Previous molecular dynamics (MD) studies have investigated how some of the structural features evolve as a function of the cooling rate.\cite{vollmayrCoolingrateEffectsAmorphous1996,laneCoolingRateStress2015} Because slow cooling is computationally demanding, these studies are usually based on a handful of glass structures and those quenched at fastest rates are often more akin to the inherent structures of liquid (i.e., local minima from the liquid region of PEL) than glass. To overcome this issue, we apply replica exchange molecular dynamics (REMD) to anneal silica structures below the glass transition temperature in the simulation (i.e., the fictive temperature, T\textsubscript{f}). This allows us to acquire BKS silica glass structures that are more stable than those attainable with the cooling rates utilized in previous melt-quenching studies. We then apply several machine learning methods to predict the thermodynamic stability from multiple structural features using a dataset consisting of 24,157 inherent structures across the glass and liquid regions. The structural features most critical to the prediction are identified, which sheds light on the structure-stability relationship in vitreous silica and demonstrates how machine learning methods can be used to resolve complex multivariate problems in glassy systems.

\section{Methods}
\subsection{Preparation of glass structures}\label{MD_method}
To prepare the dataset for machine learning, silica glass structures with different thermodynamic stabilities are generated using two MD methods. Note that we define thermodynamic stability in this study as the enthalpy of the structure under 1 bar pressure. The first method utilizes the melt-quenching technique. Here, silica melts are quenched linearly from 5,000 K to 300 K using five different cooling rates of 0.01, 0.1, 1, 10 and 100 K/ps. The constant pressure---constant temperature (NPT) ensemble with 1 bar pressure is applied throughout the process. Snapshots are taken every 1 ps from each trajectory and instantaneously relaxed to 0 K to obtain their "inherent structures". The inherent structures reflect the the local energy minima on the potential energy landscape and provide the basis for analyzing the structure-stability relationship. 

While the melt-quenching technique is able to generate structures with a large range of stability, it is inefficient in sampling PEL once the simulation temperature is below the glass transition. To overcome this issue, we employ replica exchange molecular dynamics (REMD).\cite{yamamotoReplicaexchangeMolecularDynamics2000} This method establishes a series of independent MD simulations, or replicas, running simultaneously at different temperatures. At given time intervals (every 10 ps in this work), attempts are made to exchange replicas at neighboring temperatures based on an exchange probability. To ensure the entire simulation remain at equilibrium in the thermodynamic ensemble, the exchange probability has the form of\cite{hukushimaExchangeMonteCarlo1996}
$$P_{n,n+1}=\min(1,e^{(\beta_n-\beta_{n+1})(H_n-H_{n+1})})$$
where $n$ denotes the $n^{th}$ simulation replica, $H$ denotes the enthalpy, $ \beta=(k_{B}T)^{-1} $, $k_{B}$ is the Boltzmann constant, and $T$ is the temperature. If the exchange is accepted, the two replicas are immediately adjusted to the new temperatures by rescaling the atom velocities. Due to this exchange mechanism, replicas running at the lower temperatures have a probability to cross over energy barriers at higher temperatures and sample the PEL. In this work, we run replicas in the NPT ensemble at 20 temperatures between 1,809 and 2,650 K and take trajectory snapshots every 100 ps. The inherent structures of these snapshots are then obtained by energy minimization. The temperature range for the simulation is selected to cover the fictive temperature of silica identified in the melt-quenching simulations ($\sim$2,500 K). For simulation efficiency, the 20 temperatures are distributed to yield similar exchange acceptance rates between all the neighboring temperatures.

All simulations in this work are performed with GROMACS.\cite{abrahamGROMACSHighPerformance2015} The systems contain 1,512 Si atoms and 3,024 O atoms in cubic simulation boxes. The BKS potential is employed for all the simulations.\cite{vanbeestForceFieldsSilicas1990} While the BKS potential has been shown to reproduce experimentally observed structural features in silica, it has a known issue that the glass density is sensitive to the potential cutoffs.\cite{vollmayrCoolingrateEffectsAmorphous1996,cowenForceFieldsMolecular2015,sundararamanMechanicalPropertiesSilica2016} In this study, due to a large cutoff of 8.5 {\AA} used for both short and long range potentials, the simulated densities (around 2.5 g/cm$^3$, as detailed in Results) are higher than that from experiments (2.2 g/cm$^3$). Nonetheless, the trend of density change with temperature agrees with those from previous studies using a cutoff of 5.5 {\AA}.\cite{vollmayrCoolingrateEffectsAmorphous1996} 
We employ a time step of 1 fs, the velocity-rescale algorithm for temperature coupling, and the Berendsen algorithm for pressure coupling.\cite{berendsenGROMACSMessagepassingParallel1995} For energy minimization, we use an MD-based approach to instantaneously quench the structure to 0 K under 1 bar pressure, instead of optimization methods like conjugate gradient. This approach allows us to overcome discontinuities due to the force field cutoffs. The enthalpy of the inherent structure, $e_{\mathrm{IS}}$, is calculated after convergence at 0 K and further used for machine learning.
 
 In total, 24,157 inherent structures are obtained from the MD simulations. Their structural features and inherent enthalpy are used for machine learning. The structures as well as the machine learning dataset are available for download from the Materials Data Facility.\cite{https://doi.org/10.18126/uffm-o07g}

\subsection{Machine learning methods}\label{ML_method}
 The inputs to machine learning consist of 70 numeric structural features of silica glass, including values related to the mass density, the  coordination defects, the total and partial pair distribution functions (PDFs), the total and partial structure factors, the rings of different sizes ($n$=3--10), the distributions of torsion angles and Si-O-Si bond angles. A detailed list of all the features is provided in Appendix. These features are computed for all the inherent structures generated from the MD simulations. For the features determined based on connectivity, e.g., coordination defects and rings, a cutoff of 2.0 Å is used to establish the neighbor list. This cutoff is selected based on the location of the minimum after the first peak in the Si-O partial PDF. Details of these structural features are discussed in the Results section. Because the features considered here can have values in very different scales, we normalize each input value by $X'=(X-X_{\mathrm{min}})/(X_{\mathrm{max}}-X_{\mathrm{min}})$, i.e., rescaling it to 0--1, before applying machine learning methods. The output of the machine learning models is the enthalpy of the inherent glass structure, $e_{\mathrm{IS}}$. 85\% of the dataset are randomly taken as the training group and the remaining 15\% the test group. The training data are further learned in different machine learning models with a leave-15\%-out multiple folder cross validation. The test data are hidden from the training process and used to evaluate the accuracy of the machine learning predictions. 

To investigate the structure-energy relationship, we apply both linear and non-linear machine learning algorithms. For linear algorithms, we utilize the least absolute shrinkage and selection operator (Lasso) method, which has been used previously to resolve complex descriptor-property relationships in materials.\cite{tibshiraniRegressionShrinkageSelection1996,ghiringhelliBigDataMaterials2015} In this work, it has a the loss function of $||Xw-e_{\mathrm{IS}}||_2^2+\alpha||w||_1$, where $X$ is the matrix of structural features, $w$ is the vector of weights, and $e_{\mathrm{IS}}$ is the vector of target, the inherent BKS enthalpies. $\alpha$ is a regularization parameter balancing the prediction error and the number of non-zero weights, and it is determined independently from the training of $w$. $\alpha$ in this work is selected corresponding to the first plateau in the plot of the mean square error vs. $\alpha$, which ultimately results in 31 non-zero weights. A leave-15\%-out 10-folder cross validation is applied to the model training to avoid overfitting. We also apply the Tikhonov regularization (also called Ridge regression), which has a loss function of $||Xw-e_{\mathrm{IS}}||_2^2+\alpha||w||_2^2$, to confirm our conclusions regarding feature significance from the Lasso model.\cite{hoerlRidgeRegressionBiased1970} At similar levels of prediction accuracy, the Lasso model tends to minimize the number of non-zero weights, while Ridge tends to minimize the absolute values of the weights. The linear machine learning methods are implemented within the Scikit-learn package for python3.\cite{pedregosaScikitlearnMachineLearning2011} A constant intercept for all structures is included in the linear models. 

For non-linear regression, we apply the algorithms of the artificial neural network (ANN) and the decision tree. In the ANN model, we implement 1 hidden layer with 20 nodes and 1 output layer, selected based on both accuracy and efficiency. The transfer function in the hidden layer is a tansig function and the transfer function in the output layer is a linear function. A leave-15\%-out 6-folder cross validation is applied with the Levenberg-Marquardt backpropagation training method, which is implemented within MATLAB.\cite{haganTrainingFeedforwardNetworks1994} The settings for the decision tree can be found in Supplementary Material (SM).\cite{SM}

\section{Results}
\subsection{$e_{\mathrm{IS}}$ distribution}\label{eis}

\begin{figure}
  \includegraphics[width=\linewidth]{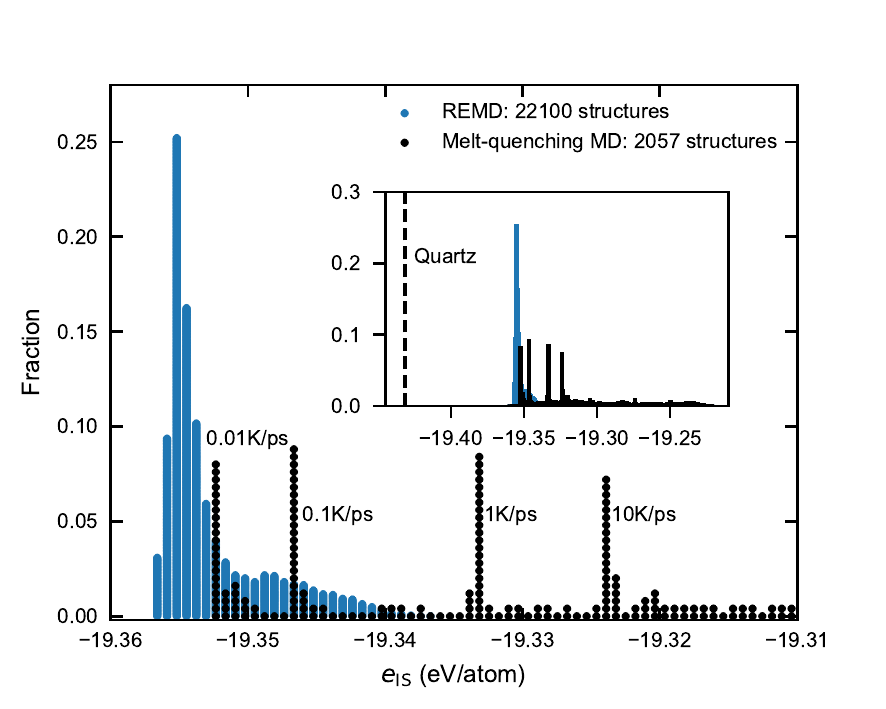}
  \caption{$e_{\mathrm{IS}}$ distributions of the inherent structures from REMD and melt-quenching MD simulations, respectively normalized by the total number of structures from each technique. The number of structures from REMD are more than 10 times of that from the melt-quenching MD. The black peaks from the melt-quenching simulations correspond to the inherent structures collected below T\textsubscript{f} that have similar inherent enthalpies. The inset shows the distribution in a broader range with the $e_{\mathrm{IS}}$ of $\alpha$-quartz noted by the dashed line.}
  \label{fig:eIS}
\end{figure}

Figure \ref{fig:eIS} shows the $e_{\mathrm{IS}}$ distribution of the structures generated from REMD (22,100 structures) and melt-quenching MD (2,057 structures). It can be seen that a majority of the REMD structures have lower $e_{\mathrm{IS}}$ than the most stable melt-quenched structure (obtained with 0.01 K/ps cooling rate). This is due to the fact that the REMD simulations in this study anneal silica below the fictive temperature (T\textsubscript{f}$\sim$2,500 K). In addition, inherent structures collected below T\textsubscript{f} from one melt-quenching simulation (corresponding to the sharp black peaks in Fig. \ref{fig:eIS}) are very similar (the root of mean square distance \textless 0.03 \AA). In contrast, structures from REMD can be substantially different even when they have similar $e_{\mathrm{IS}}$, suggesting they are sampled from different regions of the PEL. Overall, the two techniques complement each other for the purpose of this study--the melt-quenching MD can quickly produce structures with relatively low stability (e.g., inherent structures of the silica liquid) and REMD is more efficient in generating a large number of relatively more stable glass structures. 

In the following sections, we first examine the correlation between single structural feature and $e_{\mathrm{IS}}$, and then apply different machine learning methods to investigate the overall structure-stability relationship. It is worth emphasizing that structures with $e_{\mathrm{IS}}<-19.35$ eV/atom represents 75\% of the entire dataset. While it appears as a tiny region in the plots shown below, structures from this region of high stability are likely more representative of real silica glass and carry substantial weights during machine learning.

\subsection{Coordination defects}\label{defects}

The ideal silica glass structure under ambient pressure consists of a network of corner-sharing tetrahedra, in which case the structure only contains fourfold coordinated Si ($\mathrm{Si^{IV}}$) and twofold coordinated O ($\mathrm{O^{II}}$). Silica structures generated from MD simulations can have some Si and O with different coordination numbers, which are considered as coordination defects. The most common coordination defects are $\mathrm{O^{III}}$ and $\mathrm{Si^{V}}$. Because the occurrence of other coordination defects is relatively rare, the numbers of $\mathrm{O^{III}}$ and $\mathrm{Si^{V}}$ in one structure are approximately equal so the stoichiometry is satisfied. As the glass structure becomes more stable, the overall population of the defects is expected to decrease.  

\begin{figure}
  \includegraphics[width=\linewidth]{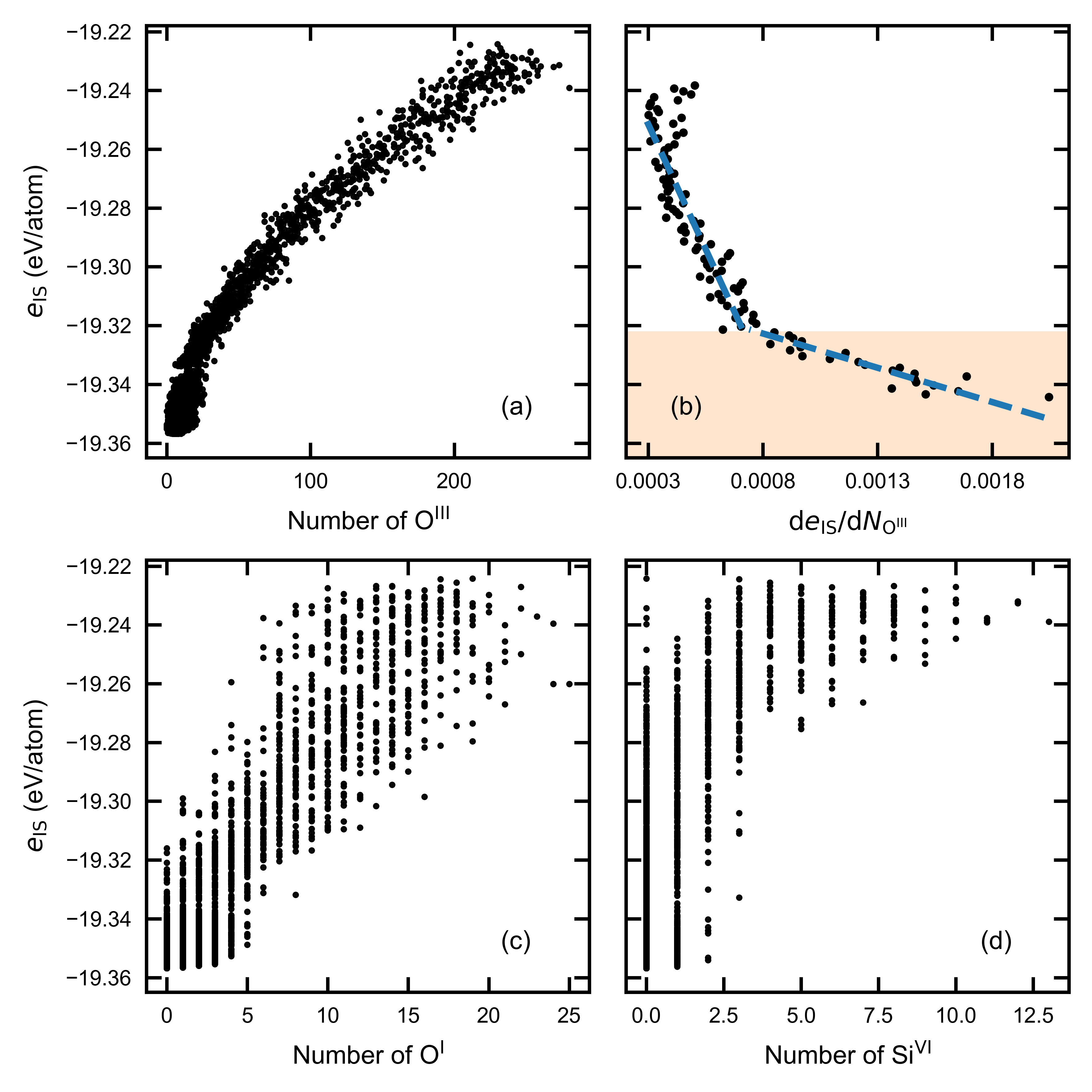}
  \caption{Plots of $e_{\mathrm{IS}}$ vs. the numbers of coordination defects: (a) $\mathrm{O^{III}}$; (c) $\mathrm{O^I}$; (d) $\mathrm{Si^{VI}}$. (b) shows the slope of $e_{\mathrm{IS}}$ with respect to the number of $\mathrm{O^{III}}$ from (a), calculated using moving average. A transition can be observed at $e_{\mathrm{IS}}$ corresponding to a fictive temperature of around 3,100 K. The dashed lines and the shaded region in (b) are added to guide the eye.}
  \label{fig:defect}
\end{figure}

Figure \ref{fig:defect} shows the relationships between $e_{\mathrm{IS}}$ and the numbers of coordination defects. It is clear that there is a correlation between $e_{\mathrm{IS}}$ and the $\mathrm{O^{III}}$ population. As expected, more stable structures have fewer $\mathrm{O^{III}}$ overall. However, notable variations in $e_{\mathrm{IS}}$ exist among structures with similar $\mathrm{O^{III}}$ populations. Furthermore, the correlation with $e_{\mathrm{IS}}$ becomes non-linear as $\mathrm{O^{III}}$ approaches zero. A change in the slope can be seen in Fig. \ref{fig:defect}b, which occurs around $e_{\mathrm{IS}}$ corresponding to liquid silica at $\sim$3,100 K. Below this transition, a large number of relatively stable structures have a near zero population of $\mathrm{O^{III}}$, while still exhibiting significant differences in $e_{\mathrm{IS}}$. 

A similar trend in $e_{\mathrm{IS}}$ is observed for $\mathrm{O^I}$, as shown in Fig. \ref{fig:defect}c. However, $\mathrm{O^I}$ concentration is very low in all the structures, and therefore the relationship is significantly scattered. More scattering is observed in the case of the even less common $\mathrm{Si^{VI}}$. In general, structures with $\mathrm{O^I}$ and $\mathrm{Si^{VI}}$ tend to also have large numbers of other defects. Consequently, high concentrations of $\mathrm{O^I}$ and $\mathrm{Si^{VI}}$ suggest high eIS. However, low concentrations of $\mathrm{O^I}$ and $\mathrm{Si^{VI}}$ do not preclude the structure from having other more common defects, such as $\mathrm{O^{III}}$. As such, the correlation with $e_{\mathrm{IS}}$ is rather weak for $\mathrm{O^I}$ or $\mathrm{Si^{VI}}$. Coordination defects other than the four types discussed above are rare in this study (e.g., number of $\mathrm{Si^{III}}<$5 in all the structures). 

It is worth noting that, edge-sharing units (or 2-member rings) are associated with $\mathrm{O^{III}}$ in silica, and therefore its population shows a similar correlation with $e_{\mathrm{IS}}$. The populations of all these coordination defects are considered as descriptors during machine learning.

\subsection{Si-O-Si bond angles and torsion angles}\label{angles}

\begin{figure}
  \includegraphics[width=\linewidth]{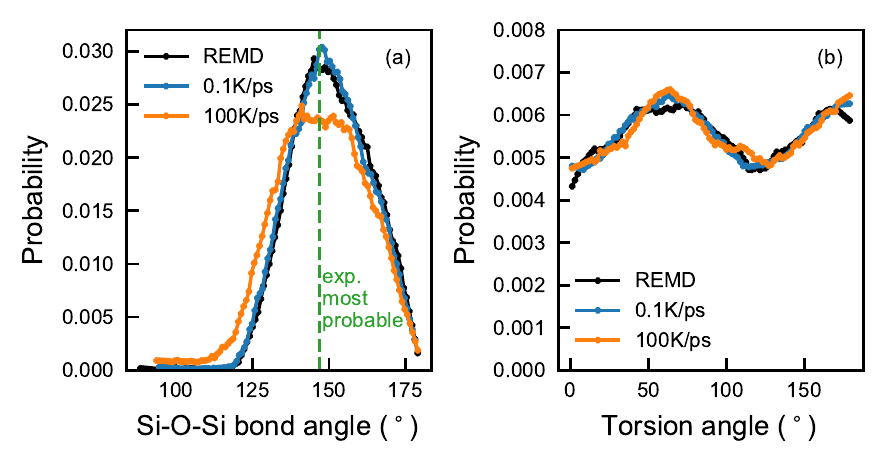}
  \caption{Distribution of (a) the Si-O-Si bond angles and (b) the torsion angles in silica structures generated by REMD, melt-quenching MD with cooling rates of 0.1 and 100 K/ps. The $e_{\mathrm{IS}}$ of the 100 K/ps and 0.1 K/ps melt-quenched structures are 0.052 and 0.010 eV/atom above the REMD structure, respectively. The plots have been smoothed for clarity. The dashed line in (a) shows the data from XRD experiments.\cite{poulsenAmorphousSilicaStudied1995,neuefeindBondAngleDistribution1996}}
  \label{fig:angle1}
\end{figure}

Next, we look at two angles defining the inter-tetrahedral geometry, i.e., the Si-O-Si bond angle and the torsion angle.\cite{wrightNeutronScatteringVitreous1994,yuanSiSiBond2003} Figure \ref{fig:angle1} shows the distributions of the two angles in three example glass structures, respectively from 100 K/ps melt-quenching MD, 0.1 K/ps melt-quenching MD, and REMD. The REMD structure shown in this analysis is the most stable structure obtained in this study. The $e_{\mathrm{IS}}$ of the two melt-quenched structures are 0.052 and 0.010 eV/atom above the $e_{\mathrm{IS}}$ of the REMD structure, respectively. 

As shown in Fig. \ref{fig:angle1}a, the Si-O-Si bond angle distribution in the simulated glasses shows a broad peak between 120$^{\circ}$ and 180$^{\circ}$. The most probable Si-O-Si bond angles (e.g., 148$^{\circ}$ for REMD) in our simulations are in agreement with values reported in experiments.\cite{yuanSiSiBond2003,koharaIntermediaterangeOrderVitreous2005} The full width at high maximum (FWHM) is 35$^{\circ}$ for the REMD structure. This value is larger than those from experiments,\cite{yuanSiSiBond2003,koharaIntermediaterangeOrderVitreous2005} but consistent with previous MD simulations using the BKS potential.\cite{yuRevisitingSilicaReaxFF2016} The glass made with the 100 K/ps cooling rate has a lower peak height, a more symmetrical shape, and a larger peak width, comparing with the other two structures. It also has a noticeable amount of angles smaller than 100$^{\circ}$ that are associated with coordination defects. For the two more stable structures, differences are more subtle. The REMD structure has a slightly higher asymmetry with more angles larger than 160$^{\circ}$ and fewer angles smaller than 140$^{\circ}$. On the other hand, no significant difference is observed among the torsion angle distributions in these three structures, all showing two peaks respectively located around 65$^{\circ}$ and 180$^{\circ}$, as shown in Fig. \ref{fig:angle1}b.

\begin{figure}
  \includegraphics[width=\linewidth]{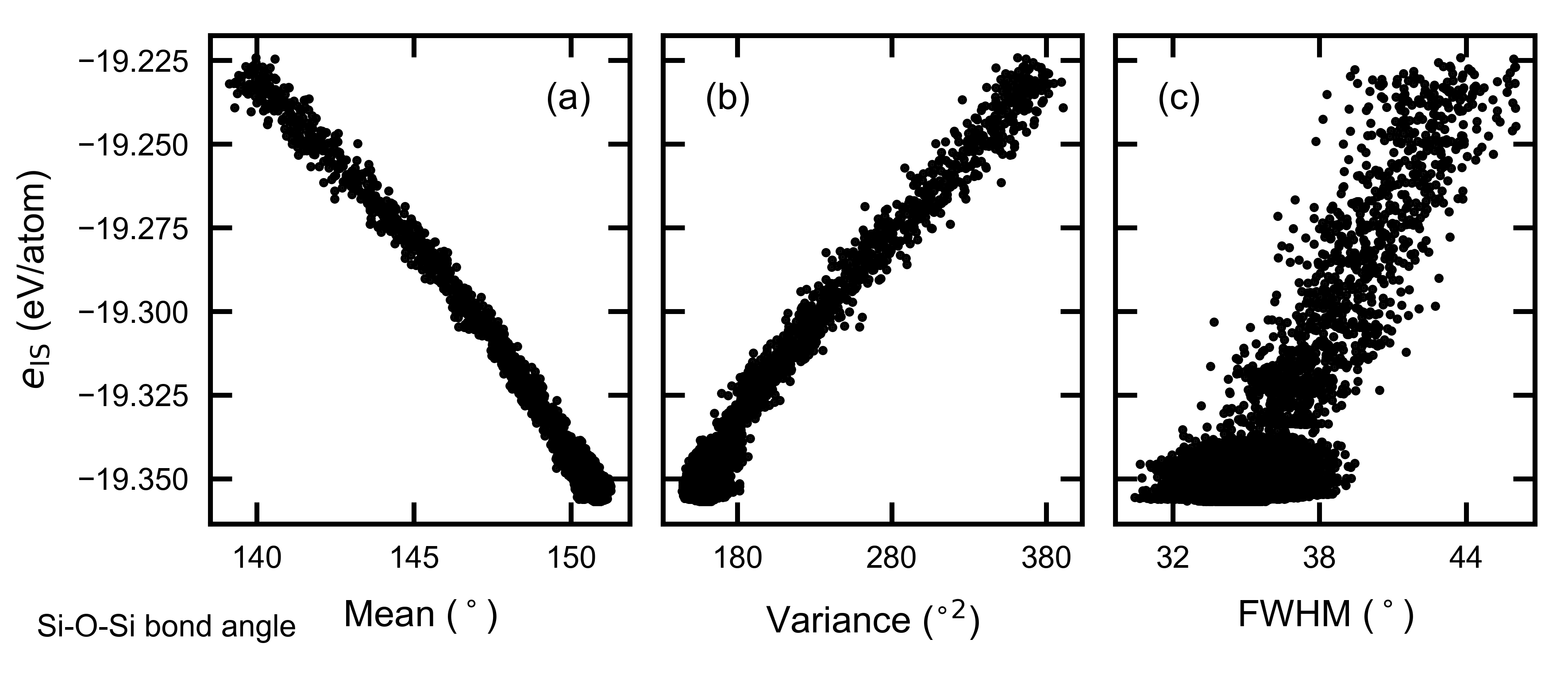}
  \caption{Plots of $e_{\mathrm{IS}}$ vs. (a) the mean, (b) the variance of the Si-O-Si bond angles, and (c) the full width at half maximum (FWHM) of the main peak in the Si-O-Si bond angle distribution.}
  \label{fig:angle2}
\end{figure}

Based on the observation above, we plot $e_{\mathrm{IS}}$ versus the mean, the variance, and the FWHM of the Si-O-Si bond angle distribution in Fig. \ref{fig:angle2}. It should be noted that, the asymmetry in the Si-O-Si bond angle distribution is reflected in the mean value, i.e., a larger mean value corresponds to a lower symmetry. From Fig. \ref{fig:angle2}, $e_{\mathrm{IS}}$ decreases with the angle mean and increases with the variance. $e_{\mathrm{IS}}$ also shows a general increase with the FWHM, although the relationship is much more scattered. These features, as well as similar properties of the torsion angle distribution (the location, the width, and the prominence of the first peak, as well as the maximum height and the variance), are used as descriptors for machine learning.

\subsection{Pair distribution functions and structure factors}

\begin{figure}
  \includegraphics[width=\linewidth]{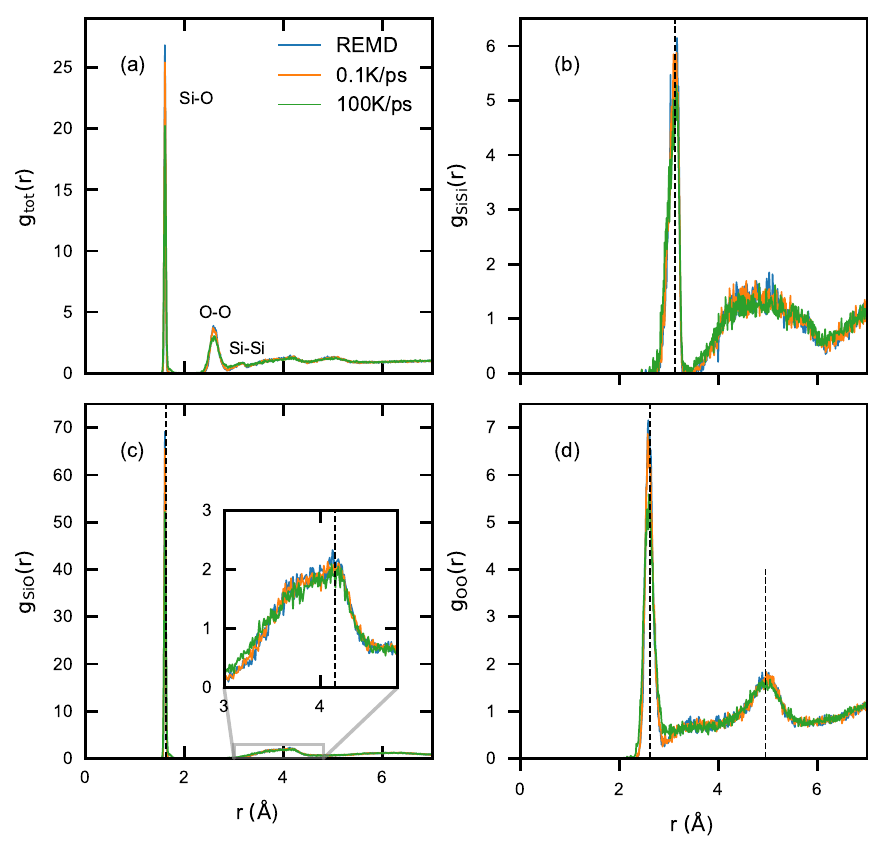}
  \caption{(a) Simulated total pair distribution functions (PDF) of the three silica structures generated by REMD, melt-quenching MD with cooling rates of 0.1 and 100 K/ps. (b-d) show the partial PDFs. The vertical dashed lines show the peaks positions identified by experiments.\cite{mozziStructureVitreousSilica1969a,grimleyNeutronScatteringVitreous1990}}
  \label{fig:pdf}
\end{figure}

Figure \ref{fig:pdf}a shows the total pair distribution functions (PDF) for the three structures, computed using partial PDFs (Fig. \ref{fig:pdf}b-d) and the neutron scattering factors.\cite{searsNeutronScatteringLengths1992} The peak locations in the PDFs show good agreements with experimental data.\cite{mozziStructureVitreousSilica1969a,grimleyNeutronScatteringVitreous1990} The three peaks below 3.3 {\AA} correspond to the distances between the nearest Si-O, O-O, and Si-Si, as can be seen from the partial PDFs. The three structures have similar peak locations, suggesting that the average bond lengths do not change significantly with $e_{\mathrm{IS}}$. In fact, none of the peak locations below 7 {\AA} show significant difference among the simulated structures. The most noticeable difference among the three structures lies in the heights of the first two peaks in the total PDF. Similar observation was reported in previous studies of the cooling rate effect in silica.\cite{vollmayrCoolingrateEffectsAmorphous1996}  Here, the peak height decreases (e.g., by about 30\% in the case of the first peak) from the most stable, REMD structure, to the least stable, 100K/ps structure. This decrease is related to a higher concentration of coordinate defects and also reflects a larger variation in bond lengths in less stable structures. 

\begin{figure}
  \includegraphics{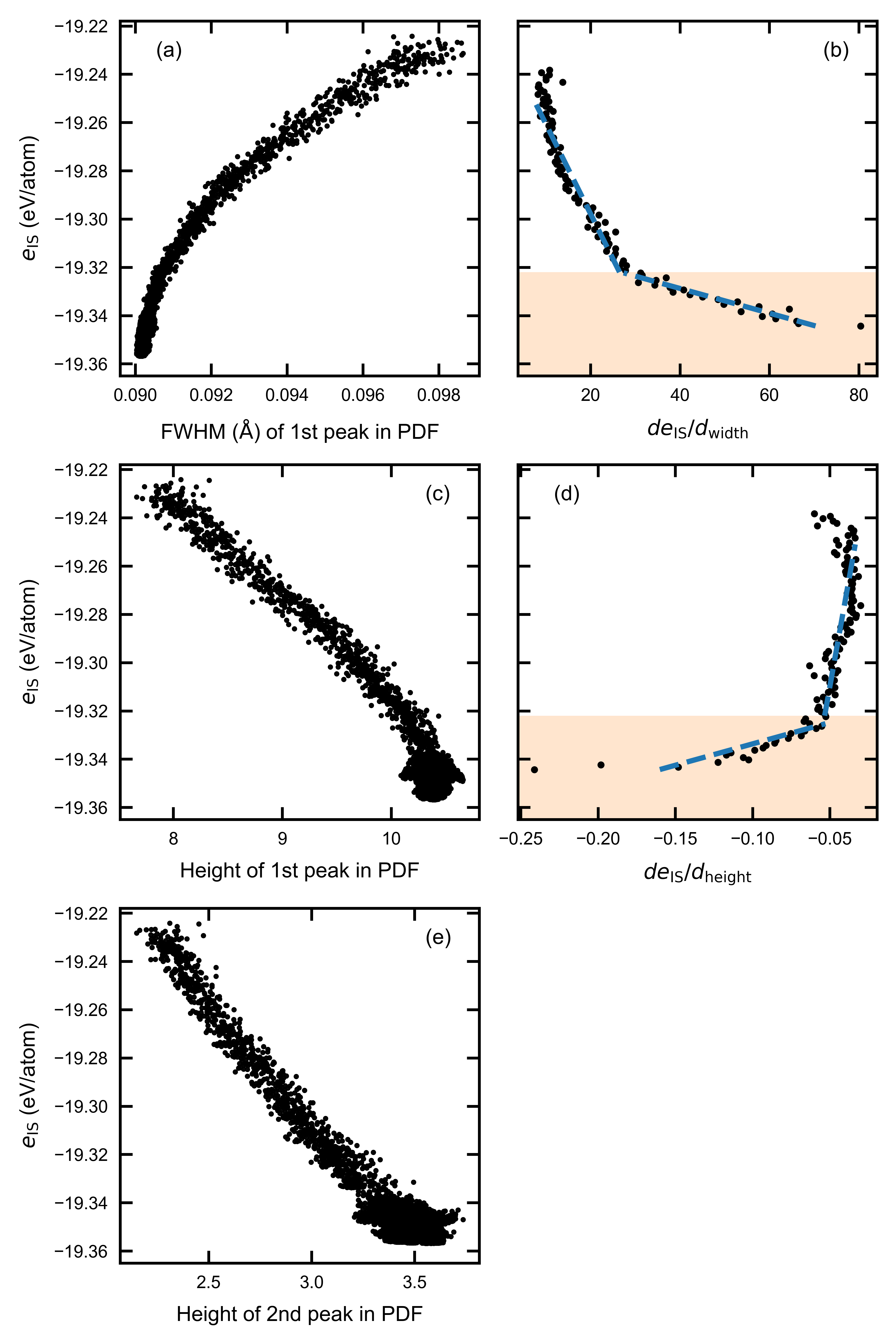}
  \caption{Plots of $e_{\mathrm{IS}}$ vs. (a) the FWHM of the 1\textsuperscript{st} peak in the total PDF, (c) the height of the 1\textsuperscript{st} peak in the total PDF, and (e) the  height of the 2\textsuperscript{nd} peak in the total PDF. We smooth the total PDFs using a window length of 0.1 {\AA} before extracting the peak information.(b) and (d) show the derivatives of $e_{\mathrm{IS}}$ with the width and the height of the 1\textsuperscript{st} peak in PDF, i.e., the slopes in (a) and (c), respectively. A transition can be observed here in (b) and (d), similar to Fig. \ref{fig:defect}b.}
  \label{fig:pdf2}
\end{figure}

Based on the observation above, we select the location, the height, and the FWHM of the first two peaks in the total and partial PDFs as descriptors for machine learning. In Fig. \ref{fig:pdf2}, both the peak height and the FWHM show strong correlations with $e_{\mathrm{IS}}$. In particular, the correlation with the FWHM in Fig. \ref{fig:pdf2}a shows the least scattering among all the features investigated in this study. However, as with coordination defects, the features of the first peak (the Si-O bond length) show a clear transition at $e_{\mathrm{IS}}$ corresponding to liquid silica at $\sim$3,100 K (see Fig. \ref{fig:pdf2}b and \ref{fig:pdf2}d). Below the transition, many stable structures have similar feature values but different $e_{\mathrm{IS}}$, rendering the features less predictive of $e_{\mathrm{IS}}$. On the other hand, the correlation with the height of the second peak appears more linear throughout the entire $e_{\mathrm{IS}}$ range. 

\begin{figure}
  \includegraphics[width=\linewidth]{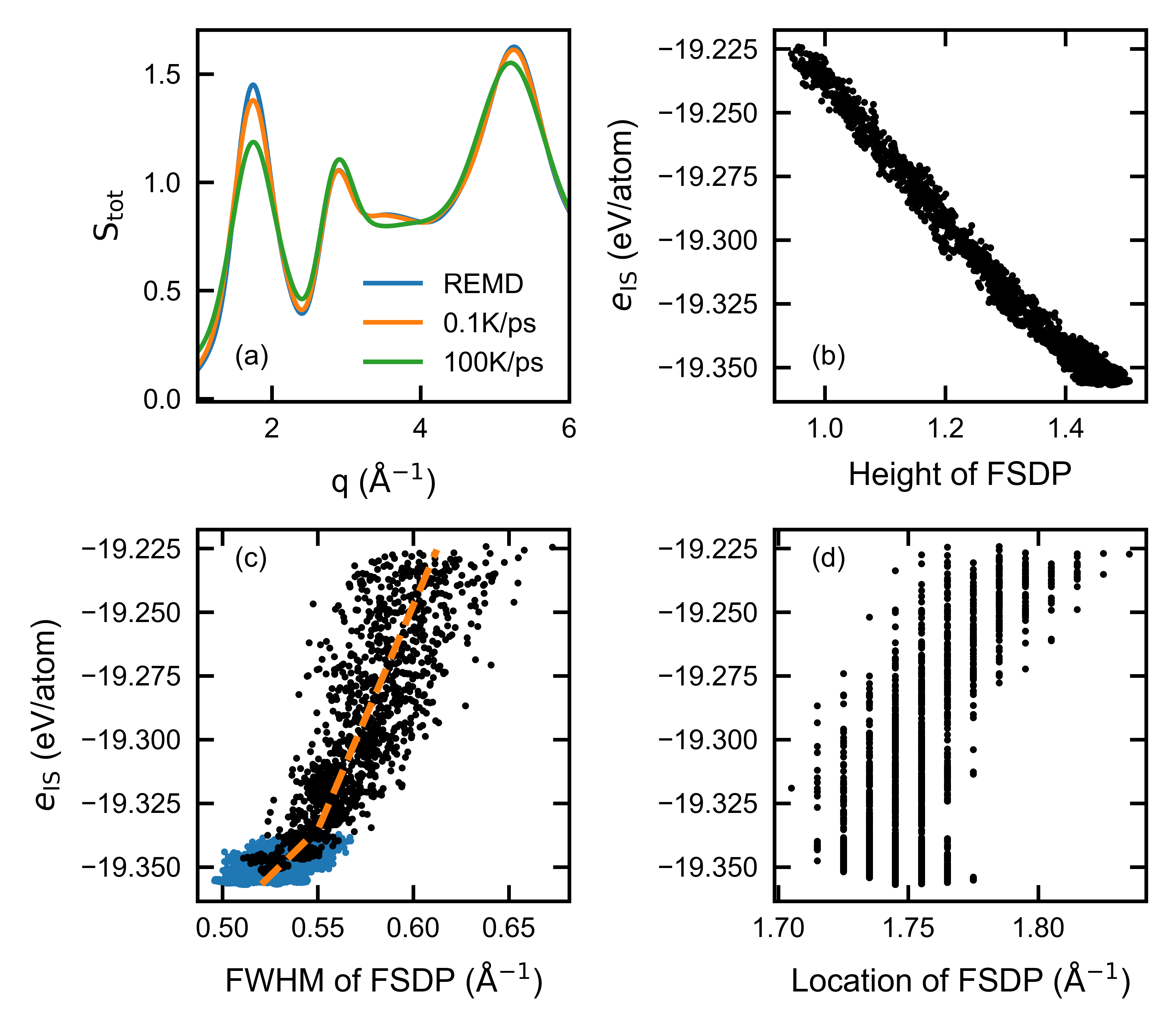}
  \caption{(a) Simulated structure factors of the three silica structures generated by REMD, melt-quenching MD with cooling rates of 0.1 and 100 K/ps. (b-d) Plots of $e_{\mathrm{IS}}$ vs. (b) the height, (c) the FWHM, and (d) the location of the FSDP. The dashed line in (c), which is a guide for the eye, shows a transition in the trend.}
  \label{fig:sq}
\end{figure}

We also compute the structure factors S(q) by Fourier transform of the corresponding PDFs, shown in Fig. \ref{fig:sq}a. Here, we focus on the first two peaks, the first and second sharp diffraction peaks (FSDP and SSDP), in the S(q) as representations of medium-range order. The most evident difference among the three structures is in the height of FSDP, which increases with stability, e.g., by about 20\% in the total S(q) from the 100K/ps structure to the REMD structure.  This is confirmed by a strong correlation between the height of FSDP and $e_{\mathrm{IS}}$ shown in Fig. \ref{fig:sq}b. Unlike with the short-range features, the correlation here is roughly linear over the entire range of $e_{\mathrm{IS}}$. We also identify the correlation between $e_{\mathrm{IS}}$ and the FWHM of FSDP (see in Fig. \ref{fig:sq}c). There appears to be a transition between less stable liquid structures from melt-quenching simulations to more stable REMD structures. The FWHM, which is inversely related to a coherent length in the real space, decreases faster in the stable structures (the blue region in Fig. \ref{fig:sq}c). However, for the most stable structures, the real space length associated with the FWHM ($2\pi/\mathrm{FWHM}$=11.5--12.5 \AA)\cite{mossRandomPackingStructural1985a,lucovskySymmetryDeterminedMedium2009} is approaching the cutoff used for computing the PDFs, i.e., half of the simulation box length ($\sim19.5$ {\AA}). Therefore, more data from larger systems are needed to confirm this observation on the coherent length. Nonetheless, we include the location, height, and FWHM of the first two peaks in the total and partial S(q)s as descriptors for machine learning. 

\subsection{Ring structures}

\begin{figure}
  \includegraphics[width=\linewidth]{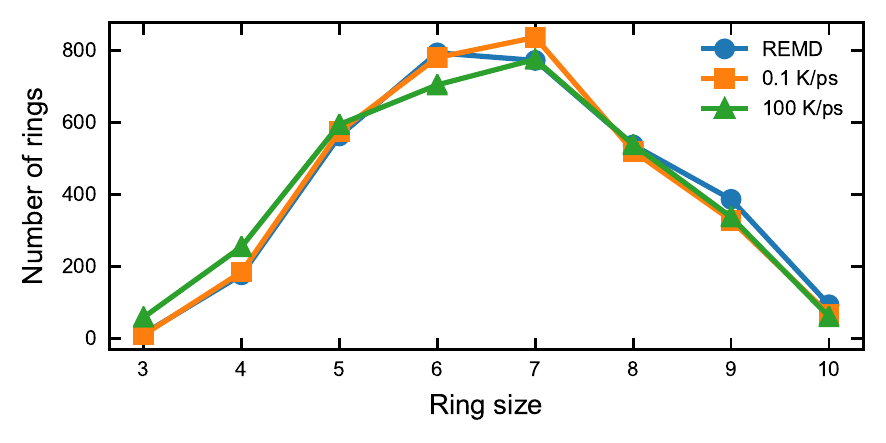}
  \caption{Ring size distributions (i.e., the numbers of 3- to 10-member rings) in silica glasses generated by REMD, and melt-quenching MD with cooling rates of 0.1 and 100 K/ps. }
  \label{fig:ring}
\end{figure}

The topology of the corner-sharing tetrahedral network in silica can be characterized by the Si-O-Si-O… ring structures. The network in crystalline quartz contains 6-member primitive rings, i.e., rings consisting of 6 Si and 6 O atoms that do not contain smaller rings. In silica glass, the ring size varies from 2 to above 10. The 2-member ring represents edge-sharing units and are discussed in Sec. \ref{defects}. We hereby evaluate rings with sizes greater than 2. 

\begin{figure}
  \includegraphics[width=\linewidth]{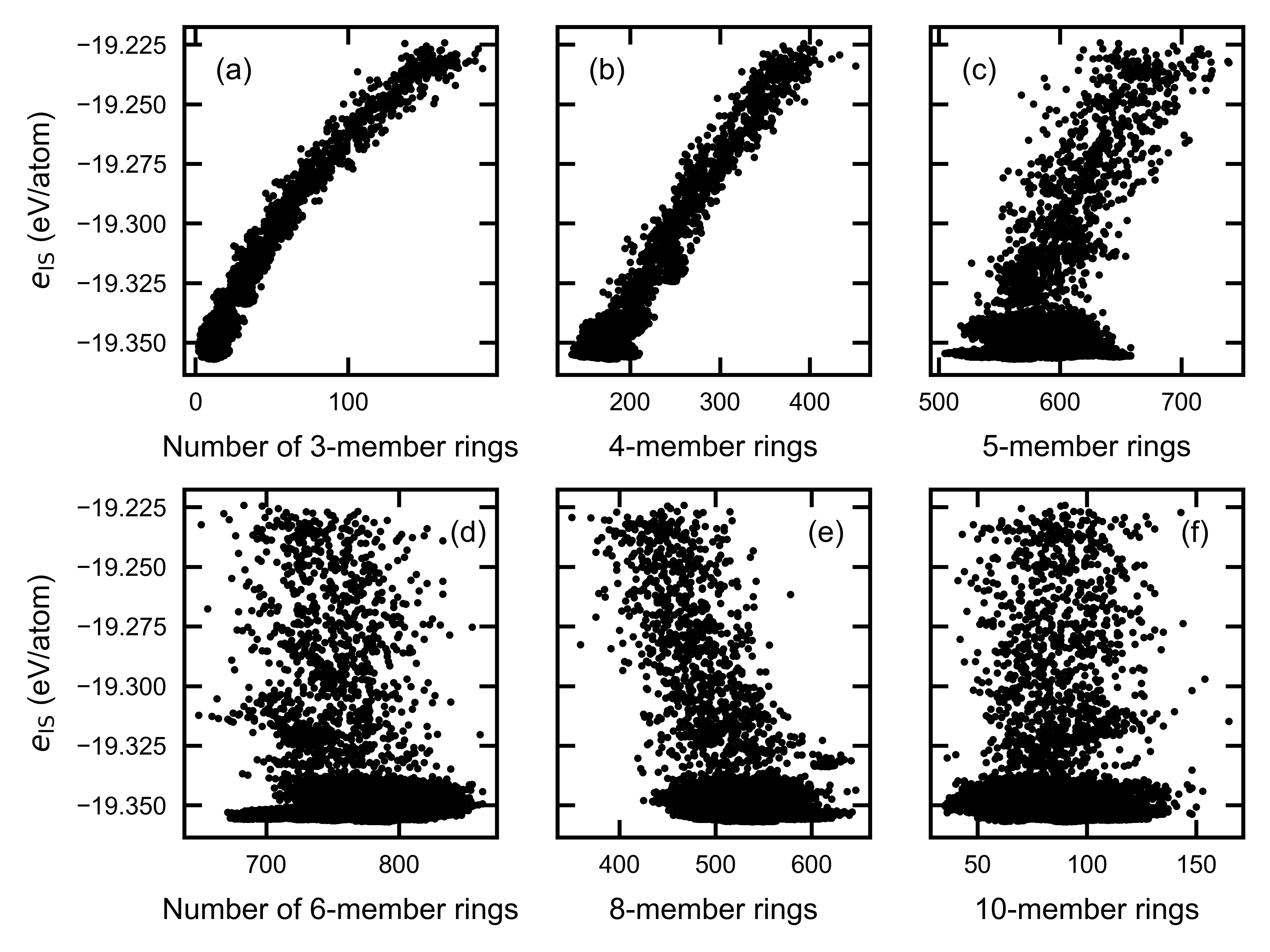}
  \caption{Plots of $e_{\mathrm{IS}}$ vs. the numbers of primitive rings: 3,4,5,6,8,10-member rings in (a-f), respectively.}
  \label{fig:ring2}
\end{figure}

Figure \ref{fig:ring} shows the distributions of primitive rings for the three glass structures. All three structures show a broad distribution between 3- and 10- member rings with a peak located at 6 or 7. The difference among the three structures is subtle -- the number of small rings (3, 4 or 5 -member rings) decreases as the structure becomes more stable. To further illustrate this, we plot the relations between $e_{\mathrm{IS}}$ and the ring populations in Fig. \ref{fig:ring2}. It can be seen that the populations of 3- and 4- member rings are inversely correlated with stability. Since these rings are associated with small Si-O-Si bond angles, this is consistent with the results shown in Sec. \ref{angles}. Significant scatterings are observed with larger rings (Fig. \ref{fig:ring2}c-f), suggesting weaker correlations. The populations of 3- to 10- member rings are all included as descriptors during machine learning.

\subsection{Density}\label{density}

\begin{figure}
  \includegraphics[width=\linewidth]{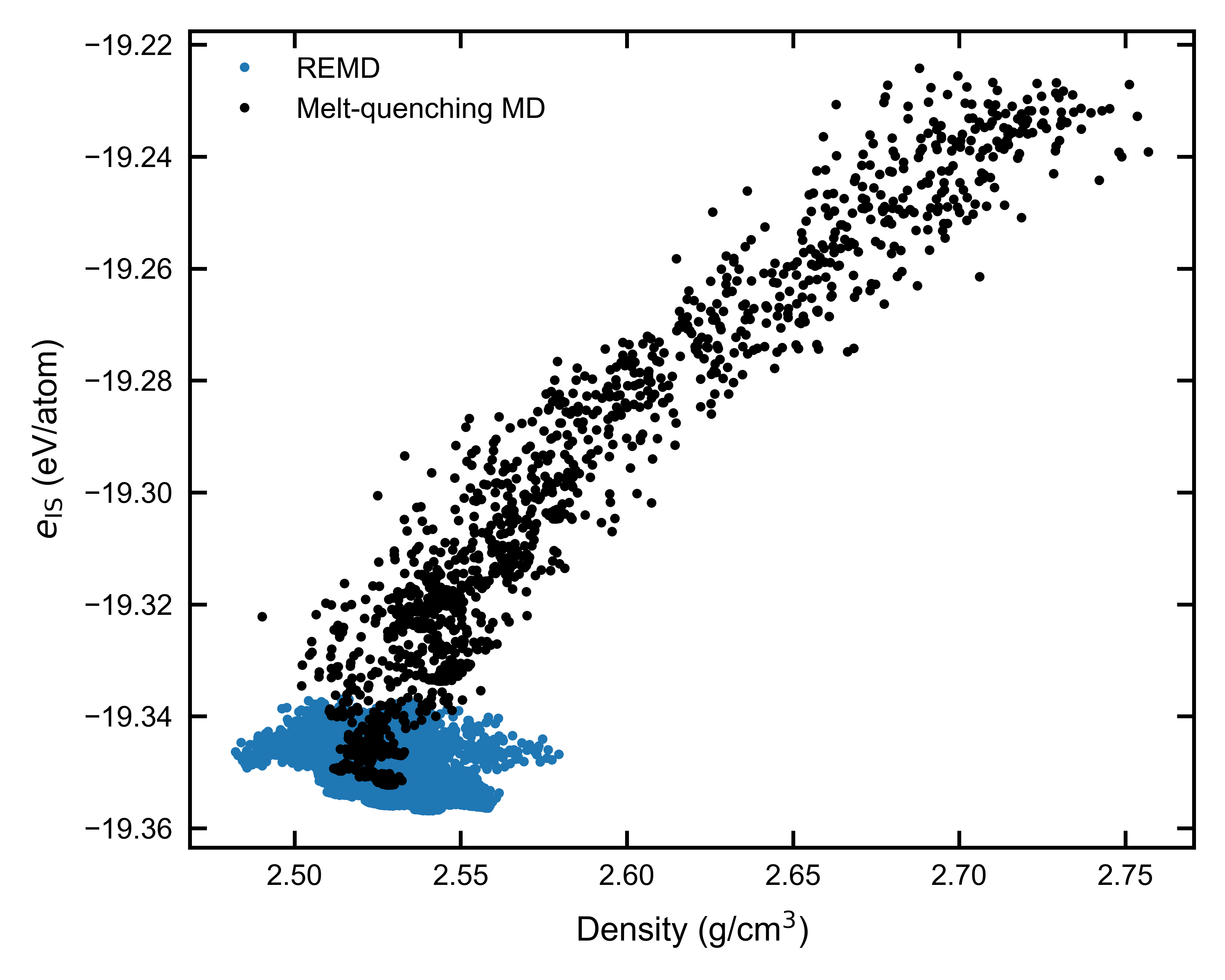}
  \caption{Plots of $e_{\mathrm{IS}}$ vs. density. The structures from REMD and melt-quenching MD are distinguished by color.} 
  \label{fig:density}
\end{figure}

Figure \ref{fig:density} shows the correlation between density and $e_{\mathrm{IS}}$. In this figure, we separate the REMD structures from the melt-quenched structures. While density appears to correlate linearly with stability, this correlation exists only for structures with relatively low stability. For these structures, a lower stability indicates a higher density. While seemingly counterintuitive, this behavior is related to the increasing populations of coordination defects as the structure becomes less stable.  Such a phenomenon has also been shown to cause abnormal thermal expansion in silica glass, i.e., the maximum of density exists at 1,820 K in experiments and ~4,500 K in simulated BKS silica.\cite{brucknerPropertiesStructureVitreous1970,vollmayrCoolingrateEffectsAmorphous1996} On the other hand, the relation between $e_{\mathrm{IS}}$ and density is less clear in the regime of more stable structures generated with REMD due to a large scatter in the data. The values of these densities are higher than the experimental value due to the choice of potential cutoffs, as discussed in Sec. \ref{MD_method}. Nonetheless, density is included as a descriptor in machine learning.

\subsection{Lasso model}

\begin{figure}
  \includegraphics[width=\linewidth]{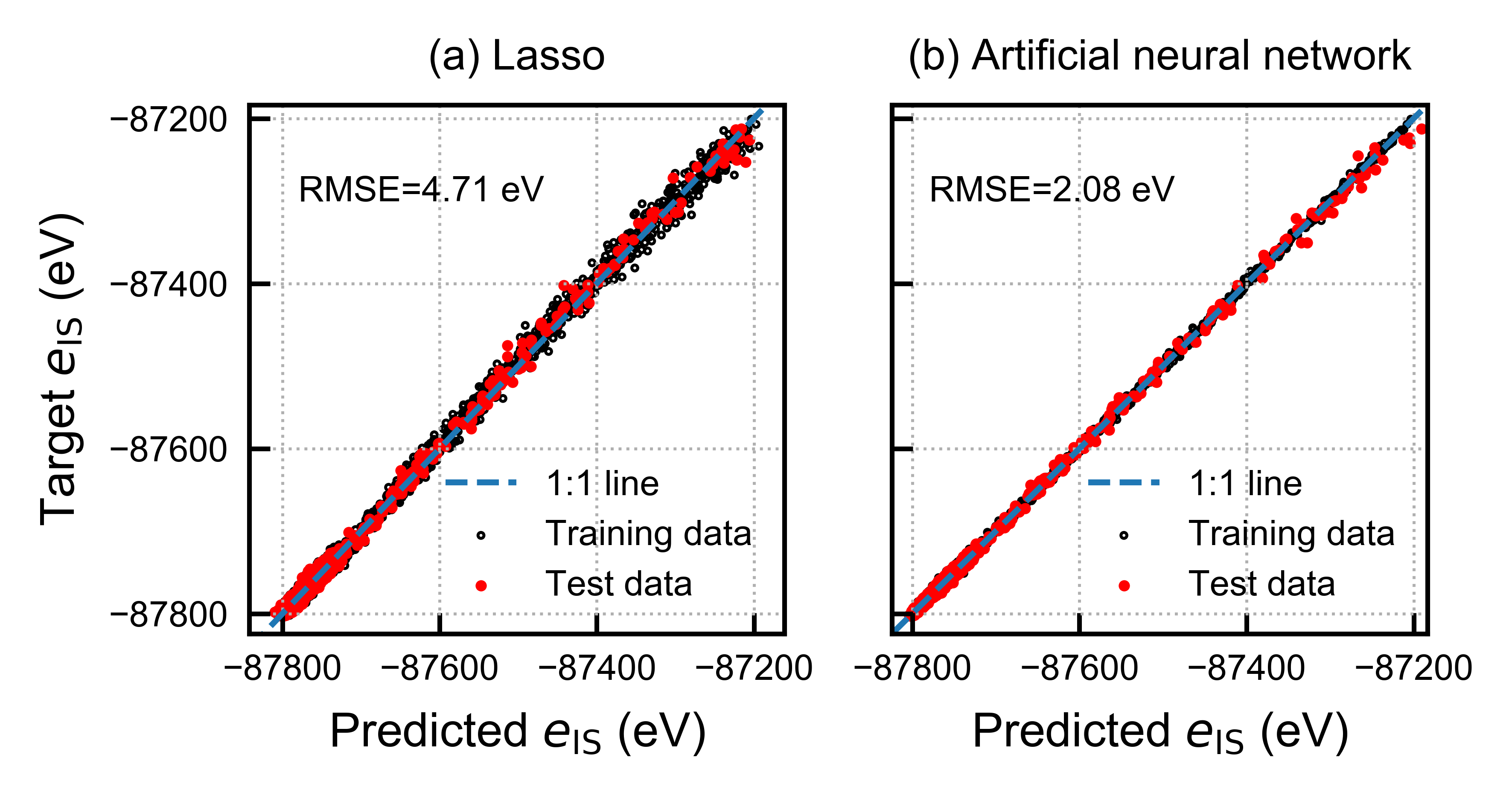}
  \caption{Performances of (a) the Lasso model, and (b) the artificial neural network model. The $e_{\mathrm{IS}}$ predicted by the models agree well with the $e_{\mathrm{IS}}$ directly calculated using the force field on both the training and testing data.}
  \label{fig:performance}
\end{figure}

Using the structural features discussed above as descriptors, we now apply machine learning to predict $e_{\mathrm{IS}}$. In this section, we train a linear model based on the Lasso regression method introduced in Sec. \ref{ML_method}, the performance of which is shown in Fig. \ref{fig:performance}a. The predicted $e_{\mathrm{IS}}$ match well with those calculated by the BKS force field. The root-mean-squared error (RMSE) for the test data is 4.71 eV, or 1.04 meV per atom. Comparing with the simple linear model (see SM for details),\cite{SM} the RMSE from Lasso is slightly higher although still sufficiently accurate. However, in the simple linear model, highly correlated features are preserved and often assigned with weights of opposite signs. In the Lasso model, features containing redundant information are removed by assigning zero weights, thanks to the regularization term $\alpha$. This allows the model to retain only those features that are critical to the prediction. The final number of non-zero weights is 31 out of 70 total features, obtained with $\alpha=0.014$. 

\begin{figure}
  \includegraphics[width=\linewidth]{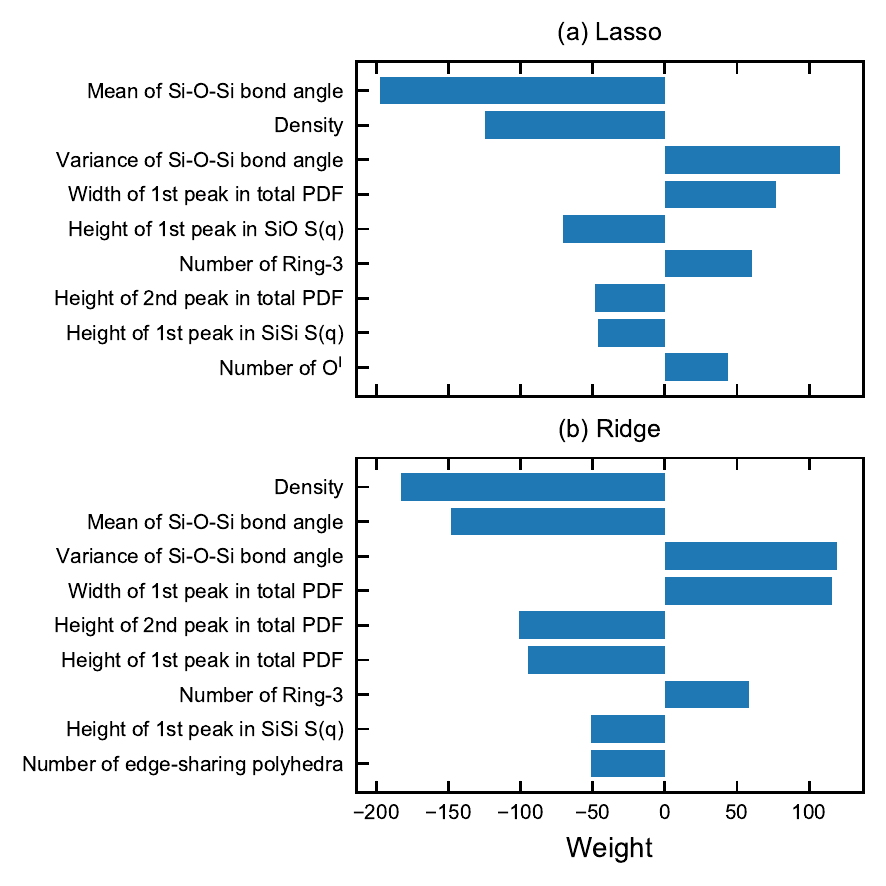}
  \caption{Features with the largest absolute values of the weights in (a) the Lasso model and (b) the Ridge model.}
  \label{fig:coeff}
\end{figure}

A benefit of the linear models is that the significance of a feature to the prediction is indicated by the absolute value of its weight, provided that the input values are normalized to the same scale. In Fig. \ref{fig:coeff}a, we list the top eight features ranked by the absolute values of the weights obtained from the Lasso regression. The signs of the weights are consistent with the correlations of these features with $e_{\mathrm{IS}}$ observed in previous sections and their corresponding Pearson correlation coefficients (see SM),\cite{SM} i.e., a positive sign corresponds to a positive overall correlation.  

We also apply another regularization algorithm, Ridge, to identify features with the largest absolute weights, shown in Fig. \ref{fig:coeff}b. The top features coincide with those from the Lasso method with a different order, which is due to the difference in the weighting algorithms. 

\subsection{Artificial neural network model}\label{ANN}

We have also employed the decision tree and the artificial neural network (ANN) methods in an attempt to better capture the non-linear associations between the structural features and $e_{\mathrm{IS}}$ observed in Sec. \ref{defects}-\ref{density}. After training, both methods show improved performance from the linear models. With the decision tree method, the five most important features are identified to be: the mean Si-O-Si bond angle, the first peak height in the Si-O S(q), the first peak width in the total PDF, the first peak height in the total S(q), and the first peak height in the Si-Si S(q). More details on the decision tree model can be found in SM.\cite{SM} Here, we focus on the ANN model which gives the best prediction in this study. The RMSE for the test data achieved from the ANN model is 2.08 eV (Fig. \ref{fig:performance}b), or 0.46 meV per atom.

While the ANN method provides excellent prediction, analyzing the underlying physical correlations from the model is non-trivial. The non-linear transfer functions used by the model do not guarantee to reflect actual physics. In addition, in contrast to linear models, weights for each node in the complex layer structures  of ANN cannot be directly used to analyze feature significance. Therefore, we perform the following tests using the strategy of input control to understand the roles of the structural features in the model. 

First, we remove one specific feature from the input dataset and retrain the model. The mean-squared error (MSE) is compared with that from the original model, which provides an indication of the irreplaceability of the feature removed. Results of this test are shown in SM.\cite{SM} In short, the three features causing the largest increases in MSE are the density, the height of the second peak in the total PDF, and the mean Si-O-Si bond angle. These three features are among the most significant features identified by the Lasso method. However, because multiple features may contain similar underlying information, the increases in MSE resulted from removing different features are found to be close and therefore cannot fully characterize the feature significance. 

\begin{figure}
  \includegraphics[width=\linewidth]{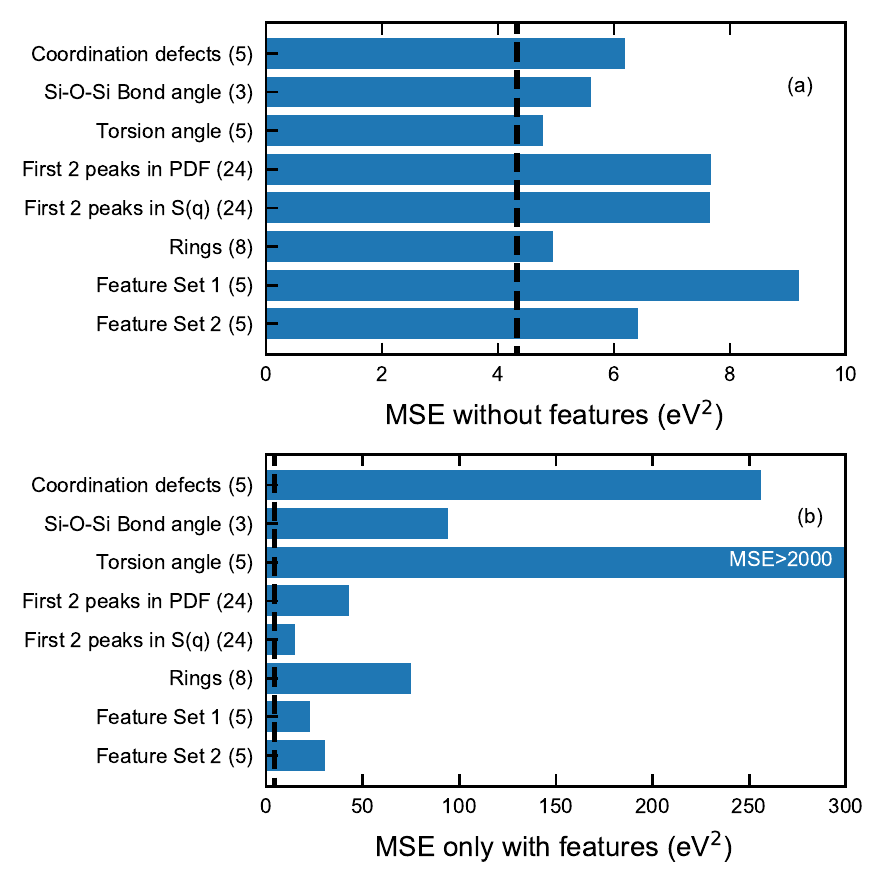}
  \caption{Mean squared errors of the ANN model trained, (a) without a feature group, and (b) only with a feature group. The number of features in each group is noted in parentheses. Feature Set 1 and Feature Set 2 are the two most important groups of features found by the Lasso method and the decision tree model, respectively. The dashed lines show the original MSE with the full set of inputs.}
  \label{fig:group}
\end{figure}

Next, we group the input structural features based on their types, e.g., all the features from S(q)s are grouped together, and we conduct two sets of tests. In the first test, we investigate the irreplaceability of a given feature group based on the MSE increase by excluding the group from the ANN training. In the second test, we train the ANN model with one feature group at a time, the MSE of which provides a direct indication of the effectiveness, or significance, of this feature group in predicting $e_{\mathrm{IS}}$. The results of the two tests are shown in Figs. \ref{fig:group}a and \ref{fig:group}b, respectively. 

Combining the results from the two tests, the role of a feature group in the ANN prediction can be evaluated. For instance, excluding torsion angle features results in a minimal increase in MSE, whereas these features by themselves give the highest MSE among all the feature groups. This result suggests that torsion angles do not contain sufficiently relevant information for predicting $e_{\mathrm{IS}}$. On the other hand, the error significantly increases when the S(q) feature group is excluded, whereas with S(q) only, the MSE is the smallest. This result suggests that the S(q) feature group contains relevant and comprehensive information for $e_{\mathrm{IS}}$ prediction. An observation similar to S(q) also applies to the PDF feature group, albeit to a slightly lesser degree. In the case of coordination defects and Si-O-Si bond angles, errors are significant in both tests, suggesting that these two groups contain useful but insufficient information. Finally, excluding ring features does not increase error but predictions based only on these features appear to be reasonable. This result suggests that ring features may contain redundant information, i.e., signatures for stability that have already been captured by other structural features. 

We also investigate two groups of features, namely the Feature Set 1 and 2, which respectively include the five most important features identified by the Lasso model and the decision tree. As shown in Figs. \ref{fig:group}a and \ref{fig:group}b, despite using only five features, both groups are found to contain irreplaceable and relevant information for $e_{\mathrm{IS}}$ prediction. In particular, Feature Set 1 shows the highest irreplaceability among all the groups tested. 

\begin{figure}
  \includegraphics[width=\linewidth]{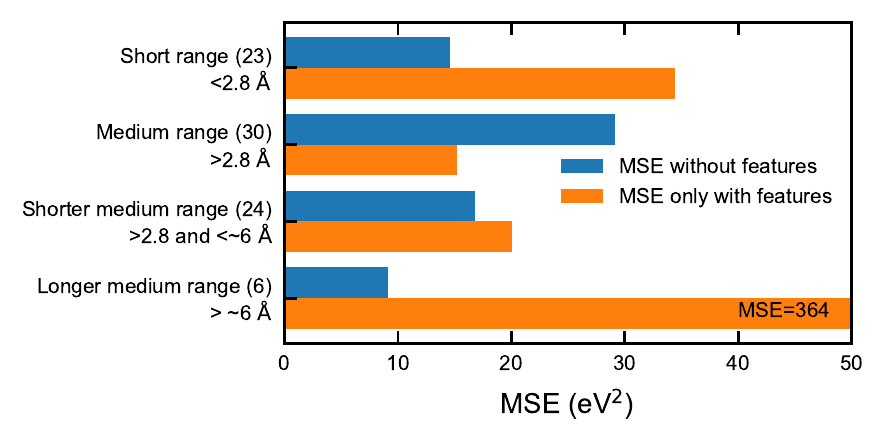}
  \caption{Mean squared errors of the ANN model trained without or only with features grouped based on their length scale. The number of features in each group is noted in parentheses. }
  \label{fig:range}
\end{figure}

The features can also be grouped by the length scale. Based on the PDFs, we group the features within 2.8 {\AA} (including the nearest Si-O and O-O neighbors) as short-range features and the rest as medium-range. The medium-range features are further divided into shorter medium-range (2.8 to $\sim$6 \AA) and longer medium-range ($> \sim$6 \AA). Details of the groups are tabulated in SM.\cite{SM} Figure \ref{fig:range} shows the MSEs obtained without or only with each length-scale-based feature group. The medium-range group shows the highest irreplaceability and the best prediction using only the group. Within the medium range, the longer medium-range features do not contain sufficiently relevant information. The short-range features, on the other hand, are useful but insufficient by themselves to predict $e_{\mathrm{IS}}$. 

\section{Discussion}
As an archetypal glass former, silica is an excellent case study for understanding the structure-energy (or structure-thermodynamic stability) relationship in glasses. Studies have shown that, despite having a near-ideal tetrahedral structure, such a relationship in silica is rather complex.\cite{shiDistinctSignatureLocal2019,vollmayrCoolingrateEffectsAmorphous1996} Previous MD simulations provided some insights into how structural features change with stability by investigating silica made with different cooling rates.\cite{vollmayrCoolingrateEffectsAmorphous1996,laneCoolingRateStress2015} However, the stability range and number of structures are very limited in these studies because sampling PEL below glass transition is difficult with the melt-quenching technique, as seen in Section \ref{eis}. In this work, we overcome this challenge by annealing silica around its fictive temperature using REMD. This approach generates a large number of glass structures from different regions of the PEL, of which a significant fraction have higher stability than glasses generated by slow melt-quenching (e.g., 0.01 K/ps). This structure library can then serve as the basis for understanding the structure-stability relationship.

As seen from the results, most of the structural features considered in this study exhibit some degrees of correlation with the inherent enthalpy $e_{\mathrm{IS}}$. As expected, stronger ordering, e.g., fewer defects and sharper peaks in various distributions, is generally associated with a more stable glass structure. In the short-range regime, the correlations between $e_{\mathrm{IS}}$ and structural features are non-linear and appear to be asymptotic. For instance, as the structures become more stable, the $\mathrm{O^{III}}$ population converges towards zero and the SiO\textsubscript{4} tetrahedra, characterized by the first Si-O peak in the PDF, become increasingly indistinguishable. In short, these short-range features are better descriptors of the inherent enthalpy for liquid than for glass. It is also worth noting that a transition in the correlations between $e_{\mathrm{IS}}$ and short-range features occurs around 3,100 K (see Figs. \ref{fig:defect} and \ref{fig:pdf2}), which coincides with the fragile-to-strong (FTS) transition in silica melt.\cite{saika-voivodFragiletostrongTransitionPolyamorphism2001,saksaengwijitOriginFragiletoStrongCrossover2004} The existence of such transition only in the short-range features supports the notion that the FTS transition in silica liquid is related to short-range defects.

The correlations between $e_{\mathrm{IS}}$ and medium range structural features are more interesting. The medium range ordering in silica can be characterized by the FSDP in the structure factor, the origin of which has been subjected to debate. Recent work attributes it to Si\--Si tetrahedra (i.e., tetrahedra consisting of five SiO\textsubscript{4} units), which exist in both silica liquid and glass.\cite{shiDistinctSignatureLocal2019} Here, we see that the FSDP position does not show a strong correlation with $e_{\mathrm{IS}}$ (see Fig. \ref{fig:sq}d), suggesting that it indeed corresponds to a common feature in vitreous silica structures of different stabilities. Meanwhile, the height of the FSDP, which indicates the degree of the medium-range order, correlates with $e_{\mathrm{IS}}$ almost linearly. Similarly, a strong correlation is also observed with another feature related to the Si\--Si structure, i.e., the inter-tetrahedral angle. On the other hand, in the ring size distributions, only small rings show strong correlation with $e_{\mathrm{IS}}$. Overall, medium-range structures that are no larger than several tetrahedra, and exhibit limited coupling with shorter or longer structures, strongly affect the thermodynamic stability of vitreous silica. Finally, based on the FWHM of the FSDP, which corresponds to the coherent length,\cite{bauchyCompositionalThresholdsAnomalies2013,micoulautAnomaliesFirstSharp2013,wangIrradiationinducedTopologicalTransition2017} there appears to be a transition that separates liquid and glass structures (Fig. \ref{fig:sq}c). This may suggest that a static length property may distinguish glass and liquid structures, although this warrants further investigation.
 
Nevertheless, all the correlations examined exhibit considerable scattering, suggesting no single feature can describe the stability to a high accuracy. On the other hand, using a combination of simple numerical descriptors extracted from these features, machine learning methods can predict $e_{\mathrm{IS}}$ with an excellent accuracy. The lowest error of 0.46 meV per atom, achieved with the ANN method, is lower than those obtained by recent works applying state-of-the-art machine learning methods to atomic systems .\cite{leeSIMPLENNEfficientPackage2019,wangDeePMDkitDeepLearning2018} It must be noted that this accuracy is achieved with the full set of 70 descriptors, the computation of which likely requires more resources than directly calculating the energy from the force field, and also that BKS silica is inherently a two-body system which may have simplified the prediction. Nonetheless, this demonstrates that the suite of structural features routinely used for characterizing disordered silica structure indeed contain sufficient information to describe its stability.  

Analyzing the significance of different features to the prediction of stability requires extra care due to potential correlations between the descriptors through unknown underlying mechanisms. In this regard, we find regularized linear methods, particularly the Lasso method, very effective. Overall, our results suggest that medium-range features, especially those with characteristic length below 6 \AA, are the most critical in predicting $e_{\mathrm{IS}}$, conforming what is observed from the correlations. On the other hand, the torsion angle is found to be the least relevant. The five most important features identified by the Lasso model are the mean Si-O-Si bond angle, the density, the FWHM of the first peak in the total PDF, the variance of the Si-O-Si bond angle, and the height of the FSDP in the Si-O S(q) (or the total S(q); both give identical accuracy, as shown in SM).\cite{SM} The significance of this set of features is also confirmed by the ANN model. Among the five features, the FWHM of the first peak in PDF describes the deviation of short-range structures from the ideal tetrahedron, the inter-tetrahedral Si-O-Si bond angle and the FSDP reflect medium-range order, and the density is a long-range, microscopic feature. This result suggests that the structural energy of vitreous silica is related to structures at different scales with medium-range features playing an outsize role. 

From these results, the five features discussed above offer a structural signature for the thermodynamic stability of vitreous silica. It will be interesting to explore whether they can be utilized by approaches such as metadynamics or reverse Monte Carlo to search for silica structures further down the PEL. In attempts to accelerate MD simulations of liquids or glasses with machine learning, the features identified in this study, particularly those from the medium range, may be considered in conjunction with short-range structural information to improve the accuracy. It may also be possible to utilize these features in experiments to evaluate the relative thermodynamic stability of different silica samples using structural characterization techniques, although additional validations with more accurate force field or \textit{ab initio} methods may be needed.

\section{Conclusion}
In this work, we apply machine learning methods to investigate the structure-stability relationship in vitreous silica using a library of 24,157 inherent structures sampled from liquid and glass BKS silica. We find both linear and non-linear machine learning methods can predict the thermodynamic stability, i.e., the enthalpy of the structure, with excellent accuracy from structural features commonly used to characterize disordered structures. The machine learning methods are also utilized to evaluate significance of the structural features to predicting the stability. The five most important features are identified to be the mean and the variance of the Si-O-Si bond angles, the density, the FWHM of the 1\textsuperscript{st} peak in the total PDF, and the height of the FSDP in the structure factor, which cover short, medium and microscopic ranges. Regarding the length scale, our study suggests that the medium-range structural features between 2.8--$\sim$6 {\AA} contain the most critical information, as compared with features from other ranges. These features may be utilized to search for silica structures with enhanced stabilities and facilitate machine learning accelerated atomistic simulations. We also find evidence that the short-range structural order is related to the fragile-to-strong transition in silica liquid while medium-range order may separate glass from liquid structures.

\begin{acknowledgments}
This research was primarily supported by NSF through the University of Wisconsin Materials Research Science and Engineering Center (DMR-1720415).This research was performed using the compute resources and assistance of the UW-Madison Center for High Throughput Computing (CHTC) in the Department of Computer Sciences. The CHTC is supported by UW-Madison, the Advanced Computing Initiative, the Wisconsin Alumni Research Foundation, the Wisconsin Institutes for Discovery, and the National Science Foundation and is an active member of the Open Science Grid, which is supported by the National Science Foundation and the U.S. Department of Energy’s Office of Science. This work also used the Extreme Science and Engineering Discovery Environment (XSEDE),\cite{townsXSEDEAcceleratingScientific2014} which is supported by National Science Foundation grant number ACI-1548562. 
\end{acknowledgments}

\appendix*
\section*{Appendix}
\begin{table}[h]
\caption{A list of all the 70 structural features used in the machine learning of this study.}
\centering
\begin{tabular}{|p{1.5in}|p{1.8in}|} 
 \hline\hline
 Type & Features \\ [0.5ex] 
 \hline
 density (1) & the mass density \\
 \hline
 Coordination defect (5) &  the numbers of $\mathrm{O^{I}}$, $\mathrm{O^{III}}$, $\mathrm{Si^{V}}$, $\mathrm{Si^{VI}}$, edge-sharing units \\
 \hline
 Pair distribution function (24) & the heights, the locations, and the FWHMs of the 1\textsuperscript{st} and 2\textsuperscript{nd} peaks in the total and the three partial PDFs \\
 \hline
 Structure factor (24) & the heights, the locations, and the FWHMs of the 1\textsuperscript{st} and 2\textsuperscript{nd} peaks in the total and the three partial S(q)s \\
 \hline 
 Rings (8) & the numbers of n-member rings (n=3-10) \\
 \hline
 Si-O-Si bond angle (3) & the mean, the variance, and the FWHM of the main peak in the distribution\\
 \hline
 Torsion angles (5) & the location, the FWHM, the prominence of the 1\textsuperscript{st} peak, the maximum height in the distribution, and the variance\\
 
 \hline\hline
\end{tabular}
\label{table:list}
\end{table}

\bibliography{test}

\end{document}

% \preprint{APS/123-QED}

\title{Supplemental Material of "Structural Signatures for Thermodynamic Stability in Vitreous Silica: Insight from Machine Learning and Molecular Dynamics Simulations"}% Force line breaks with \\

\author{Zheng Yu}
\affiliation{Department of Materials Science and Engineering, University of Wisconsin-Madison, Madison 53706, United States}

\author{Qitong Liu}
\affiliation{Department of Civil and Environmental Engineering, University of Wisconsin-Madison, Madison 53706, United States}

\author{Izabela Szlufarska}
\affiliation{Department of Materials Science and Engineering, University of Wisconsin-Madison, Madison 53706, United States}

\author{Bu Wang}%
 \email{bu.wang@wisc.edu}
\affiliation{Department of Materials Science and Engineering, University of Wisconsin-Madison, Madison 53706, United States}
\affiliation{Department of Civil and Environmental Engineering, University of Wisconsin-Madison, Madison 53706, United States}%

\maketitle

%\tableofcontents

\section*{Contents}
\begin{itemize}
    \item Relationships between the coordination defects
    \item Pearson correlation coefficients of the features with $e_\mathrm{IS}$ 
    \item Groups of the features based on length scale
    \item Extra machine learning results
\end{itemize}

\newpage
\begin{figure}[H]
  \includegraphics[width=\linewidth]{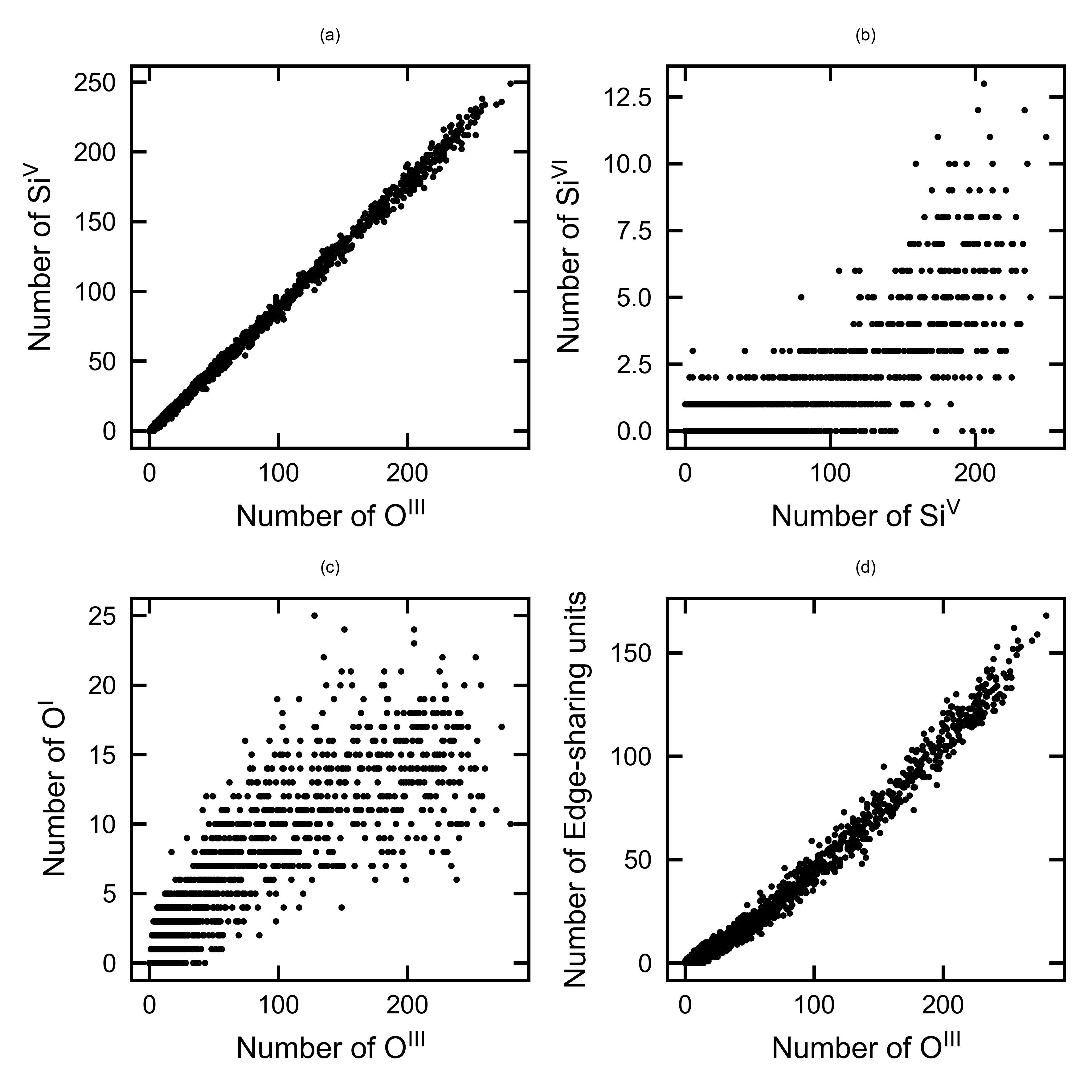}
  \caption{Relationship plots between coordination defects: (a) the number of $\mathrm{O^{III}}$ is nearly equal to the number of $\mathrm{Si^{V}}$; (b) more $\mathrm{Si^{VI}}$ correspond to more $\mathrm{Si^{V}}$ but not vice versa; (c) a significantly scattered correlation between $\mathrm{O^{III}}$ and $\mathrm{O^{I}}$; (d) a clear correlation between the numbers of edge-sharing units and $\mathrm{O^{III}}$.}
  \label{fig:cd}
\end{figure}

\begin{figure}[H]
  \includegraphics[width=\linewidth]{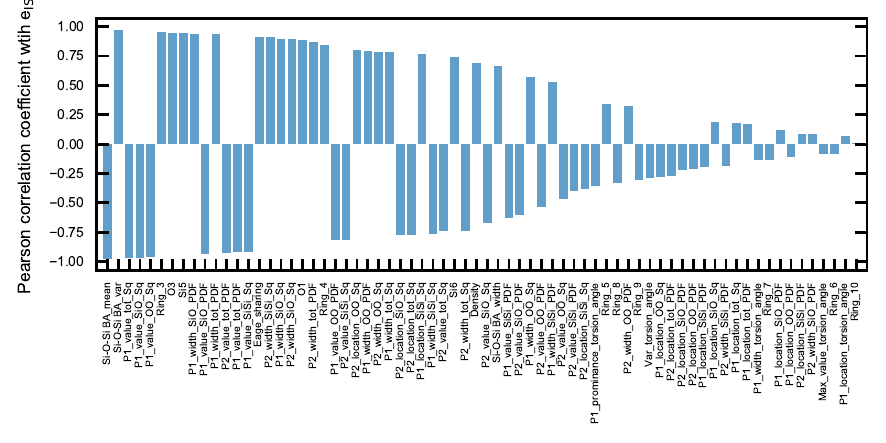}
  \caption{Pearson correlation coefficient of each feature with $e_{\mathrm{IS}}$. Pearson correlation coefficient reflects the strength of a linear correlation, in which +1 means a positive linear correlation, -1 means a negatively linear correlation, and 0 means no total linear correlation. }
  \label{fig:p_corr}
\end{figure}

\begin{table}[h]
\caption{Structural features of silica are grouped into short-range features and medium-range features based on the corresponding length scale. The boundary between short and medium range is set as the distance of the nearest O-O, such that, features within Si-O tetrahedra are considered as short-range features. The medium range is further divided into shorter medium range and longer medium range, the boundary of which corresponds to the approximate distance for symmetry fade in silica.\cite{shiDistinctSignatureLocal2019} The features from the torsion angles are excluded due to less relevance to predict $e_{\mathrm{IS}}$, and the features from the second peaks in S(q)s are excluded for their vague length ranges.} 

\medskip
\centering

\begin{tabular}{|p{2in}|p{3in}|} 
 \hline\hline
 Range (number of features) & Features \\ [0.5ex] 
 \hline
 Short $<$2.8 {\AA} (23) & the coordination defects, the first peak in the SiO PDF, the first peak in the OO PDF, the first two peaks in the total PDF \\ 
 \hline
 Medium $>$2.8 {\AA} (30) & except short range \\
 \hline
 Shorter medium \newline 2.8 $-\sim$6 {\AA} (24) & the first two peaks in SiSi PDF, the first peaks in the total and the partial S(q)s, the numbers of the rings with sizes of 3-5, the Si-O-Si bond angles \\
 \hline
 Longer medium $> \sim$6 {\AA} (6) & the density, the numbers of the rings with sizes of 6-10 \\
 \hline\hline
\end{tabular}
\label{table:1}
\end{table}

\begin{figure}[H]
  \includegraphics[width=\linewidth]{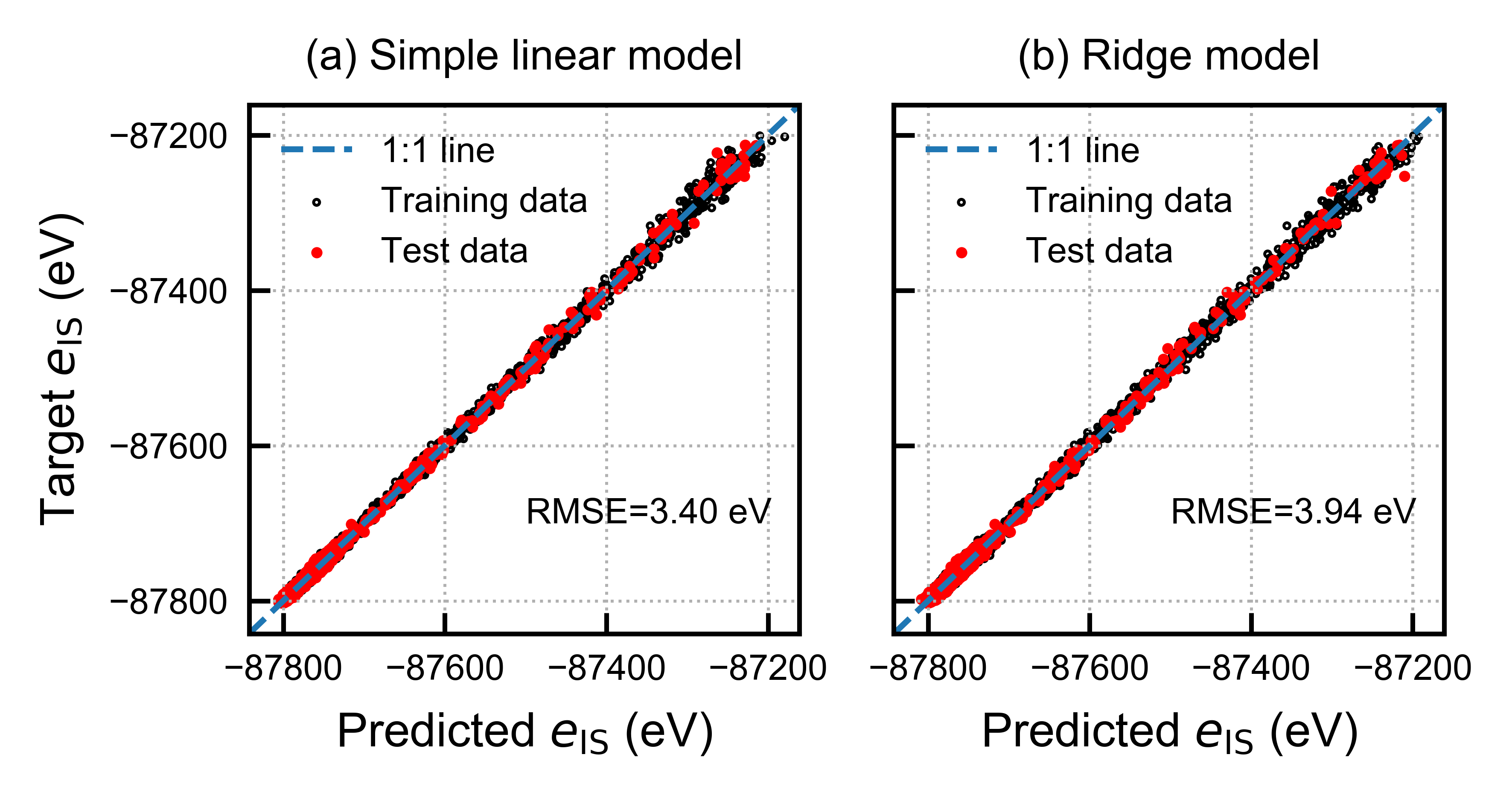}
  \caption{Performances of (a) the simple linear model, and (b) the Ridge model.}
  \label{fig:simple_linear_model}
\end{figure}

\begin{figure}[H]
  \includegraphics[width=\linewidth]{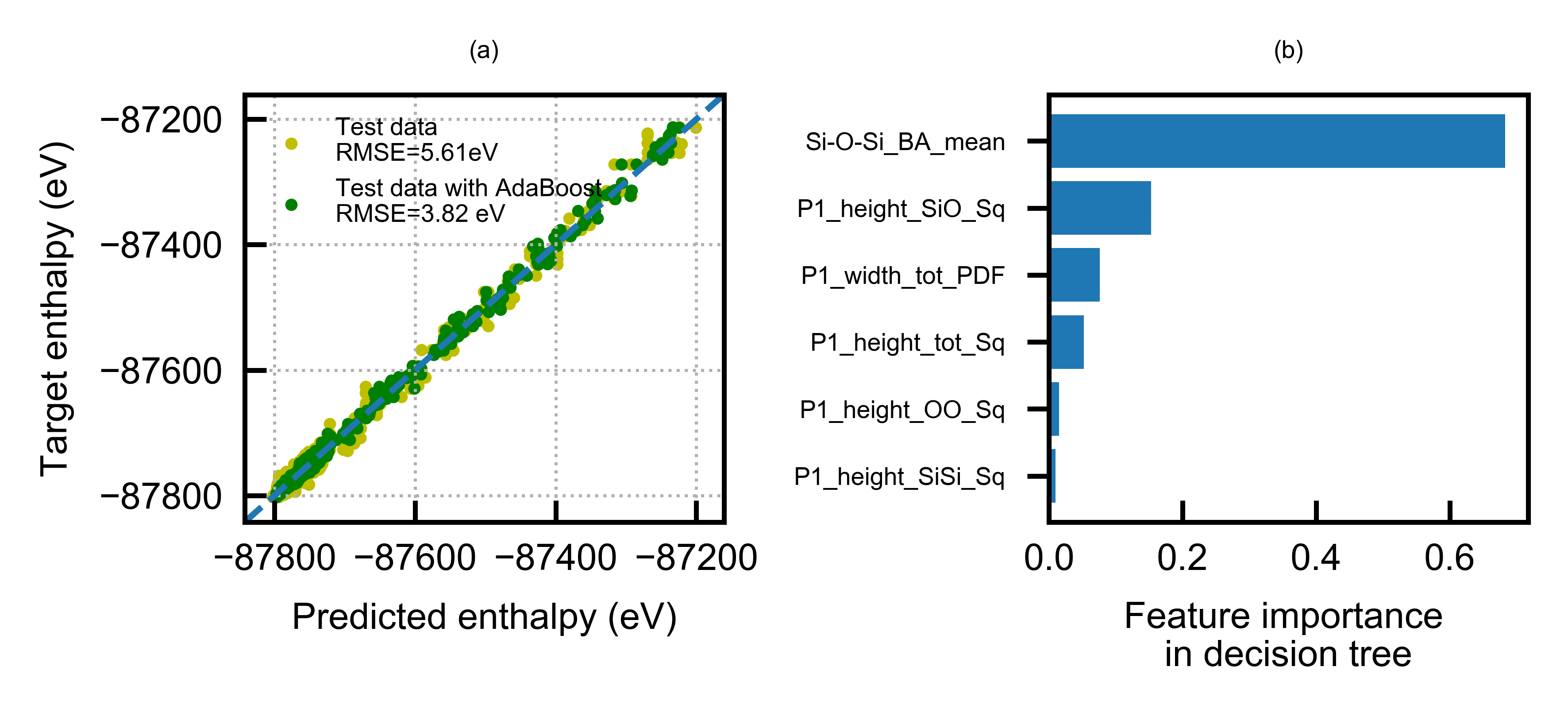}
  \caption{(a) Performance of the decision tree model (max\_depth=10) without or with AdaBoost algorithm, respectively. The RMSE of the decision tree model with AdaBoost is slightly smaller than that of the Lasso model but larger than that of the ANN model. (b) The most important features in the decision tree model with AdaBoost. The top five features are considered in the ANN feature importance analysis as Feature set 2.}
  \label{fig:decision tree}
\end{figure}

\begin{figure}[H]
  \includegraphics[width=\linewidth]{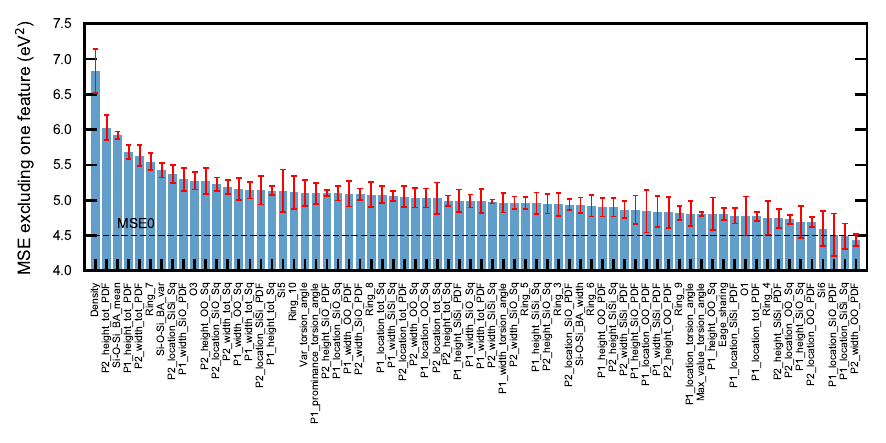}
  \caption{Mean squared error (MSE) of 70 ANN models which alternatively excludes one structural feature from training, as mentioned in Sec. IIIH in the main text. A large increase of MSE means more irreplaceability of the removed feature to predict stability. The three most irreplaceable features are found to be density, height of the second peak in total PDF, and mean of Si-O-Si bond angle. However, due to redundant information in these features, the increase of MSE are close and not enough to characterize feature significance to $e_{\mathrm{IS}}$.}
  \label{fig:ann_single}
\end{figure}

\begin{figure}[H]
  \includegraphics[width=\linewidth]{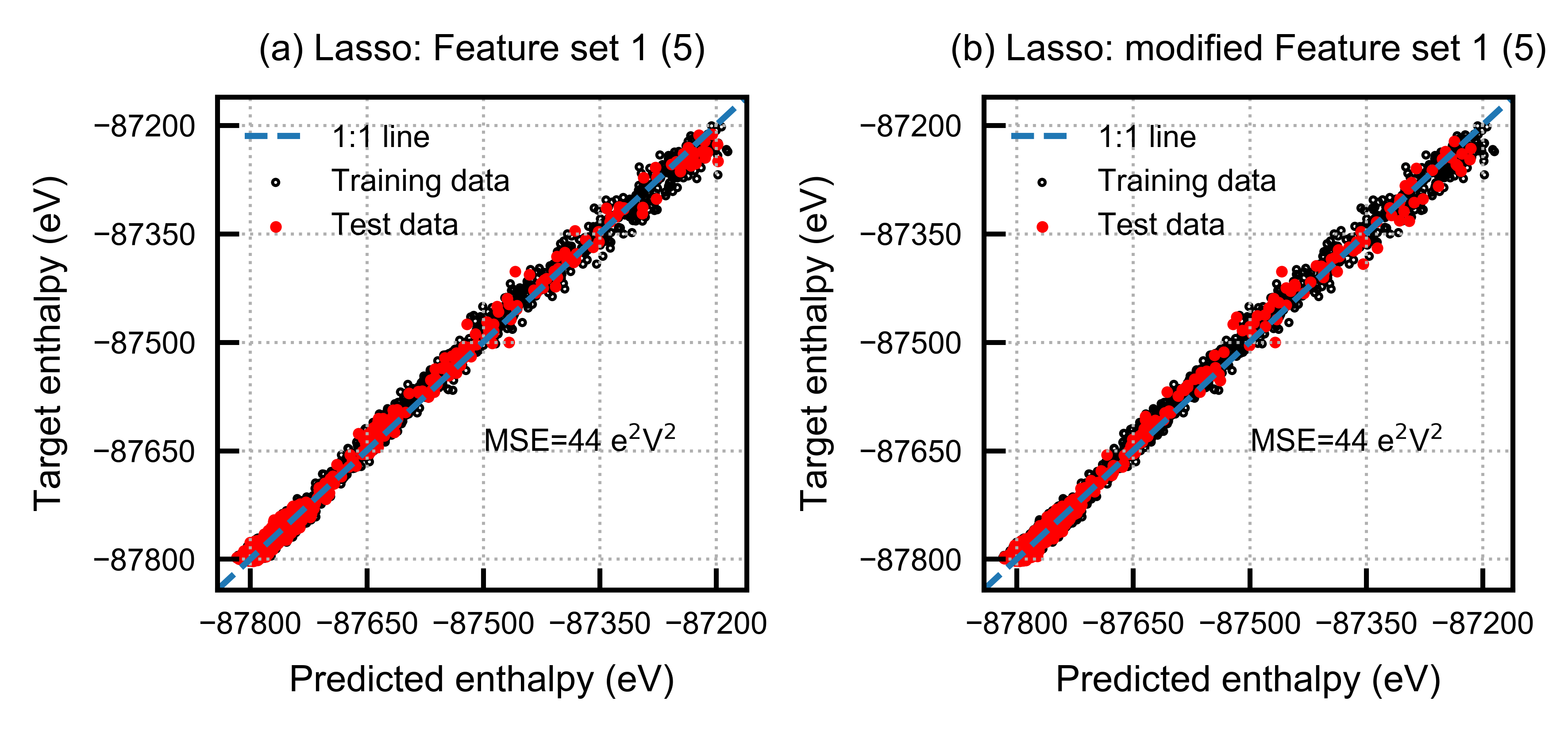}
  \caption{(a) Performances of the Lasso model using (retrain and predict) only Feature set 1, i.e., the top five predictive features from the Lasso model. The MSE increases but still remains reasonably small for $e_{\mathrm{IS}}$ prediction. (b) The same setting as (a), except that the height of the first peak in the Si-O S(q) in Feature set 1 is substituted by the height of FSDP in the total S(q). The performance does not change by this substitution. The reason we check this is that the first peak in the Si-O S(q) is a major component of FSDP, but FSDP is more convenient to measure in experiments. Because the information in the two features are mostly overlapped, a non-zero weight is only assigned to one of the two in the Lasso model. }
  \label{fig:lasso_few1}
\end{figure}

\bibliography{test}